\documentclass[letterpaper]{article} 
\usepackage{aaai2026}  
\usepackage{times}  
\usepackage{helvet}  
\usepackage{courier}  
\usepackage[hyphens]{url}  
\usepackage{graphicx} 
\usepackage{natbib}  
\usepackage{caption} 
\usepackage{algorithm}
\usepackage{algorithmic}
\usepackage{pdflscape}

\usepackage{colortbl}
\usepackage{longtable}
\usepackage{xcolor}
\usepackage{booktabs}
\usepackage{tikz}
\usepackage{rotating}
\usepackage{soul}
\definecolor{lightgray}{gray}{0.9}
\sethlcolor{lightgray}
\usepackage{newfloat}
\usepackage{listings}
\DeclareCaptionStyle{ruled}{labelfont=normalfont,labelsep=colon,strut=off} 
\floatstyle{ruled}
\newfloat{listing}{tb}{lst}{}
\floatname{listing}{Listing}
\definecolor{gtyellow}{HTML}{CDAD00}
\definecolor{gtblue}{HTML}{193670}
\definecolor{violate}{RGB}{255, 210, 180} 
\definecolor{conform}{RGB}{200, 240, 255}
\newcommand{\conform}[1]{%
  {\sethlcolor{conform}\hl{#1}}%
}

\newcommand{\violate}[1]{%
  {\sethlcolor{violate}\hl{#1}}%
}

\newcommand{\scalebar}[1]{%
  \begin{tikzpicture}[baseline=-3pt] 
    \ifdim #1 pt < 0pt
        \def\barcolor{gtyellow}
    \else
        \def\barcolor{gtblue}
    \fi
    \fill[\barcolor] (0, -2pt) rectangle (#1 * 4em, 2pt); 
  \end{tikzpicture}%
}

\usepackage{xcolor}
\newcommand{\answerYes}[1]{\textcolor{blue}{#1}} 
\newcommand{\answerNo}[1]{\textcolor{teal}{#1}} 
\newcommand{\answerNA}[1]{\textcolor{gray}{#1}}

\title{Follow the Rules (or Not): Community Norms and AI-Generated Support in Online Health Communities}
\author {
    Shravika Mittal\textsuperscript{\rm 1},
    Erin Kasson\textsuperscript{\rm 2},
    Layna Paraboschi\textsuperscript{\rm 2},
    Eleanor Laufenberg\textsuperscript{\rm 2},
    Jiawei Zhou\textsuperscript{\rm 1},
    Patricia A. Cavazos-Rehg\textsuperscript{\rm 2},
    Tanushree Mitra\textsuperscript{\rm 3},
    Munmun De Choudhury\textsuperscript{\rm 1}
}
\affiliations {
    \textsuperscript{\rm 1}Georgia Institute of Technology\\
    \textsuperscript{\rm 2}Washington University\\
    \textsuperscript{\rm 3}University of Washington\\
    smittal87@gatech.edu, erinmkasson@wustl.edu, lparaboschi@wustl.edu, e.laufenberg@wustl.edu, j.zhou@gatech.edu, pcavazos@wustl.edu, tmitra@uw.edu, munmund@gatech.edu
}
\usepackage{bibentry}

\begin{document}

\maketitle

\begin{abstract}
Generative AI (GenAI) is increasingly being integrated into the online ecosystem, including online health communities (OHCs), where people with diverse health conditions exchange social support. For example, in OHCs, support providers are beginning to share content generated, directly or indirectly, by popular GenAI-based tools. OHCs are governed by norms that define appropriate behavior when providing support. Ways in which AI-generated support interacts with these norms remain underexplored. Inappropriate conformance or outright violation can erode seekers' trust, distort decision-making, and threaten community sustenance. In this work, we examine whether (and how) AI-generated support conforms to norms, using popular opioid-use recovery subreddits as our testbed. First, we provide an inventory of norms regulating text-based support provision in OHCs. Next, using human-validated LLM judges, we assess the prevalence of AI's conformity to these norms. Finally, through an expert review, we identify risks to seekers (and OHCs) resulting from norm (non)conformity. Our analysis revealed that, while AI-generated support conforms to norms, such conformity may be inappropriate or insufficient, for example, by over- or under-validating seekers in distress. Moreover, we observed instances of outright norm violation. This work provides insights that can help moderators and OHC designers adapt existing and develop new norms to regulate AI integration, protecting both seekers and communities they rely on.
\end{abstract}


\section{Introduction}

Online health communities (OHCs), such as communities on Reddit (i.e., subreddits), are popular among people with diverse health conditions, and their caregivers, to reciprocate social support~\cite{10.1145/3461778.3462100}, including actionable advice, encouragement, and reassurance. This exchange of support is uniquely governed by norms, shared, often community-created expectations, which regulate how support is offered and sought~\cite{burnett2001small}. When upheld, these norms structure interactions in ways that allow communities to function effectively; when violated, they can undermine seekers' well-being and community sustenance~\cite{10.1145/2998181.2998277}. 

Today, generative AI (GenAI) is rapidly permeating the online ecosystem, reshaping how content is produced, shared, and consumed across platforms~\cite{techcrunchInstagramIntroduces,10.1145/3757474}. 
AI-generated content (AIGC) is becoming particularly visible within OHCs---at times transparently through platform-integrated features~\cite{techcrunchInstagramIntroduces}, and other times opaquely through undisclosed use~\cite{10.1145/3757445}. 

GenAI introduces a new form of participation within OHCs. AIGC, e.g., in response to support seekers' queries, may violate, or inappropriately conform to, community norms, posing risks to both individual members and the broader community. For example, AIGC has been shown to ``pollute'' these spaces with misleading content~\cite{Mittal_Jung_ElSherief_Mitra_DeChoudhury_2025}, violating well-established norms against sharing such information~\cite{fan2018online,winker2000guidelines}.
For many users, such as people with opioid use disorder, OHCs serve as a primary source of support~\cite{Balsamo_Bajardi_DeFrancisciMorales_Monti_Schifanella_2023}, making such non-conformance, including outright norm violations and superficial adherence, particularly consequential. It can erode trust, distort decision-making, and compromise the quality of support available.

In moments of significant change, OHCs often respond by adapting their governance structures, including introducing new norms or moderation practices to regulate participation~\cite{10.1145/2858036.2858356}. In decentralized settings, such as health-related subreddits, these decisions are frequently left to community members, who base them on assessments of whether (and how) new forms of participation adhere to shared community norms~\cite{10.1145/3757445}. With respect to AIGC, however, these assessments remain limited to norms against promoting low-quality, misleading, or derogatory content~\cite{Mittal_Jung_ElSherief_Mitra_DeChoudhury_2025,10.1145/3544548.3581318}. Consequently, it is warranted to examine AIGC's conformance to a broader set of norms, including those concerning emotional responsiveness, rapport building, and non-directive communication. 

Further, knowing whether AIGC follows these norms provides only a partial understanding. Since these norms were developed in the context of human participation, whose social expectations and capacities differ from AI, it is essential to examine how AIGC conforms to them, and where such conformity may fall short or become problematic, how violations may occur, and what risks may emerge or be exacerbated. 
In this light, our study pursues the following two research questions: 
\begin{description}
    \item [RQ1:] Does AIGC, in response to support-seeking queries, conform to norms on support provision within OHCs?
    \item [RQ2:] What problematic behaviors or risks emerge as a consequence of norm conformance or violation?
\end{description}

To answer these RQs, we begin by assessing AIGC's conformance to norms on text-based support provision within OHCs, using subreddits focused on recovery from opioid (mis)use as our testbed. First, we contribute an inventory of norms derived from a rapid review of scholarly literature and further enriched using data-driven signals---i.e., by reviewing content that received positive engagement from community members---and community-created, formalized rules. Next, using human-supervised LLM judges, we assessed conformity of GPT-4 generated responses to the identified norms (RQ1). Finally, using an open-coding approach, three authors with expertise in interpreting OHC data and clinical domain knowledge, identified risks that may emerge or escalate as a consequence of norm (non)conformity (RQ2).

We identified a range of norms, both implicitly valued by community members and explicitly codified in subreddit rules. These norms protect seekers' identity, regulate the type and delivery of support, and establish expectations for sustained engagement over time. While a substantial proportion of GPT-4 generated responses conformed to norms designed to support seekers' emotional well-being and safety, for instance, by avoiding information overload (85.24\%) and offering reassurance (60.27\%), many violated norms against providing prescriptive, directive guidance (34.99\%) and support misaligned with seekers' needs (39.99\%). Moreover, experts observed that both norm conformity and violation could, at times, lead to problematic consequences, such as stalled decision-making, cognitive overload, and reduced community engagement. Through this understanding, we offer ways in which community members and moderators can adapt existing norms and develop new ones for the emerging, hybrid human-AI setting, protecting both seekers and the support environments they rely on. 


\section{Background and Related Work}

\subsection{Norms and Online Health Communities}

OHCs, such as discussion forums and social networking sites, continue to serve as vital spaces for individuals with diverse health conditions to exchange social support~\cite{10.1145/3461778.3462100}. These communities offer logistical convenience 
 and a judgment-free environment~\cite{Balsamo_Bajardi_DeFrancisciMorales_Monti_Schifanella_2023}. 

Like other social groups, OHCs operate on norms, which are shared, often implicit, expectations that dictate acceptable behavior, thereby regulating the exchange of support~\cite{burnett2001small}. They help promote interactions that enable communities to achieve their goals~\cite{kiesler2012regulating} 
and sustain ongoing participation~\cite{arguello2006talk}. 
Norms can be structural, shaping the design of an OHC to, for example, protect support seekers' identity~\cite{graham2017failure,meng2021cancer} or they can be content-oriented, governing readability, appropriateness, or credibility of the support offered~\cite{fan2018online,winker2000guidelines,progga2023understanding}. 
In addition, norms can be broadly classified into two types~\cite{burnett2003beyond}: (1) explicit; codified in formal written documents, e.g., FAQs or community-created rules, and (2) implicit; informally understood and not necessarily documented. 

In the past, scholars have characterized norms that govern online communities, in general. For example, Fiesler et al.~\cite{Fiesler_Jiang_McCann_Frye_Brubaker_2018} conducted a mixed-methods analysis of explicit norms, i.e., formalized, community-created rules, across a variety of subreddits spanning topics such as health, entertainment, and politics. Some have also identified implicit norms, or ``hidden rules,'' by performing an empirical analysis of community-specific signals, again, across diverse subreddits---for example, by assessing linguistic patterns in content removed by moderators~\cite{10.1145/3274301} or in comments that received positive engagement from community members~\cite{park-etal-2024-valuescope}. Given that norms are context dependent~\cite{Fiesler_Jiang_McCann_Frye_Brubaker_2018}, these broad examinations flatten the nuance required to characterize norms, which specifically govern support provision within OHCs. At the same time, studies that do account for this context specificity often adopt a narrow lens~\cite{10.1145/3173574.3174240,10.1145/2556288.2557108}, with each characterizing only a limited subset of normative behaviors. To this end, our study contributes an inventory of explicit and implicit norms that govern text-based support provision within OHCs, developed through a rapid review of prior work, and enriched via data-driven signals and community metadata. This inventory can be resourceful for onboarding newcomers~\cite{lampe2014crowdsourcing}, supporting content moderators, and auditing new forms of participation, e.g., GenAI, within OHCs.

\subsection{GenAI and Online Health Communities}
With the emergence of GenAI-powered applications, various components of the online ecosystem, including online health communities, are experiencing a substantial influx of AIGC. On these platforms, members are interacting with AIGC in ways that are both transparent~\cite{techcrunchInstagramIntroduces} 
and opaque~\cite{10.1145/3757445}.

While GenAI may support OHCs in addressing problematic content~\cite{10.1145/3687030,mittal2025exposurecontentwrittenlarge}, 
AIGC also has the ability to pollute these spaces with low-quality information~\cite{10.1145/3544548.3581318}. 
If left unregulated, such content may violate the established community norms, posing risks not only to individual members but also to the overall health and sustainability of communities themselves~\cite{10.1145/2998181.2998277}.

In moments of significant technological change, communities respond by introducing new forms of governance, for example, updated norms, rules, moderation practices, or technical affordances, to navigate the shift better~\cite{10.1145/2858036.2858356}. For example, many platforms have implemented top-down changes in response to GenAI; some ban its use, 
while others encourage it or integrate GenAI-based features into their user interface~\cite{techcrunchInstagramIntroduces}. On the other hand, decentralized communities, e.g., those on Reddit, leave it up to community members to position the use of GenAI. This decentralized moderation approach allows communities to craft policies that are contextually-relevant. For example, policies are based on whether (and how) the use of a new technology conforms to communities' well-established norms and values~\cite{10.1145/3757445}. 

In the past, efforts to examine AIGC's conformity with OHC norms have largely focused on its tendency to produce low-quality information---i.e., its violation of norms against sharing such information~\cite{10.1145/3544548.3581318,Mittal_Jung_ElSherief_Mitra_DeChoudhury_2025}. In contrast, this work begins to provide a comprehensive evaluation of whether (RQ1), and in what ways (RQ2), AIGC conforms to a broader set of norms, e.g., those related to rapport building, emotional responsiveness, or non-directive communication, when providing text-based support in OHCs. This investigation can support community members in introducing new norms, rules, or policies around the use of GenAI---in contrast to relying on rigid, impractical, blanket measures such as, outright bans or mandatory disclosures---grounded in empirical evidence.

Moreover, unlike prior work that focuses on outright norm violations~\cite{park-etal-2021-detecting-community}, we also surface problematic instances of inappropriate or superficial conformance, through an expert-driven qualitative analysis of AI-driven support (RQ2). This consideration is essential, as community norms, such as the expectation to share lived experiences, presume a human support provider. As GenAI cannot authentically embody such behaviors, attempts at conformance may be counterproductive or harmful. This analysis can support communities in adapting to a hybrid environment, where both human- and AI-driven support co-exist, and guide AI developers to create responsible development workflows.

\section{Data}\label{sec:data}
To answer our research questions, we considered OHCs on Reddit, or, subreddits, which are widely used for exchanging healthcare-related support~\cite{,10.1145/3290605.3300354,Choudhury2014MentalHD}. They are also experiencing a substantial influx of content generated by, or co-authored with, AI~\cite{10.1145/3757445,10.1145/3706598.3713292}, making them a particularly relevant context for examining AI's conformance to norms governing online support provision. To appropriately scope this work, we considered subreddits discussing recovery from opioid (mis)use as our evaluation context. They are popular among people with opioid use disorder (OUD), and often serve as a primary source of social support~\cite{Balsamo_Bajardi_DeFrancisciMorales_Monti_Schifanella_2023}.

We used a publicly available, pre-processed dataset, hereafter referred to as Reddit-QA~\cite{laud2025large}. It consists of original posts, with support-seeking queries related to OUD, dated from January '18 to September '21. We purposely chose a timeframe prior to the release of ChatGPT, which popularized the presence of AIGC in OHCs. Though the dataset spans 19 subreddits, we selected a subset of 5---r/OpiatesRecovery, r/Suboxone, r/Methadone, r/Opiates, and r/OpiatesWithdrawal---that center on discussing recovery, rather than (mis)use of opioids more broadly. 
In total, we worked with $17,341$ posts, containing support-seeking queries on a variety of sub-topics such as harm reduction, mitigating withdrawal, treatment, and lifestyle changes.

Next, to study the conformance of AI-generated responses to norms on online support provision, we used GPT-4 (specifically, gpt-4-0613), which was the latest, most stable and capable text generation model available at the time of this work. We considered a temperature of $0.7$, the default for publicly accessible text-generation models or conversational agents~\cite{openai2024gpt4technicalreport}. We carefully created a prompt (see Table~\ref{tab:gpt4-prompt}) by referring to existing guidelines. 
Refer to Appendix Section ``Prompt'' for details. 

\section{Methods}
We describe our methodology for distilling an inventory of norms, both explicit and implicit, regulating text-based support provision in OHCs and for answering the RQs.

To gather norms, we first considered explicit, publicly visible, community-created rules listed in the subreddit metadata. We then identified less apparent, implicit norms using two complementary norm extraction approaches: community signal-based extraction and a scoping review of prior literature. Together, these approaches capture both implicit norms that are locally valued within the specific OHC contexts examined in this study and broader, generalizable implicit expectations surrounding online support provision that may not be evident from community signals alone.

\subsection{Explicit Norms}\label{sec:explicit_norms}
Consistent with prior work~\cite{Fiesler_Jiang_McCann_Frye_Brubaker_2018,10.1145/3706598.3713292}, we referred to the formalized community rules, listed in the metadata of the five OUD recovery subreddits, to curate a repository of explicit norms. In total, there were $42$ rules across the five subreddits. We discarded\footnote{From r/suboxone, we excluded a rule encouraging the disclosure of AI use, as assessing conformity to it would require analyzing user interactions (in practice), outside the scope of this study.} those that (1) targeted original posts (or support-seeking), (2) focused on modalities beyond text-based support (e.g., images, audio, or video), and (3) imposed community access restrictions (e.g., age limits). Two authors, with prior experience in social computing research, performed an inductive coding~\cite{thomas2003general} of the remaining rules, leading to six broad categories of explicit norms: (1) avoid sharing prescriptive guidance, (2) avoid providing off-topic responses, (3) respect anonymity, (4) combat stigma, (5) avoid financial or product-based sourcing, and (6) avoid self-promotion.

\subsection{Implicit Norms: Through Community Signals}\label{sec:comm_signal}

To begin, we inferred implicit norms from community-specific signals. We examined 50 posts drawn from the previously described Reddit-QA dataset, sampling 10 posts from each of the five subreddits.

Similar to~\citet{park-etal-2024-valuescope}, we considered community's reaction to the comments posted directly in response to the original post, or the score garnered by each (i.e., the number
of upvotes minus the number of downvotes), as an indicator of \emph{``prevailing implicit norms that govern behavior within the communities.''} From each community, we selected 10 original posts that received comments with the highest variance in score. For these posts, we then selected 2 comments; one with the highest and the other with the lowest score. We ensured that the two comments were relevant to the original post, provided (rather than sought) support, and were posted within a comparable time frame.

Two independent coders, graduate-level students with experience in clinical substance use research, reviewed each post and the corresponding high- and low-rated comments. To infer norms, each coder used an open-ended approach to interpret why a comment may have received a higher or a lower rating from the community. For example, they noted patterns in the language used, type of information or advice shared, and formatting. Each coder then synthesized their notes to generate a list of norms, suggesting community preference~\cite{park-etal-2024-valuescope}, by simultaneously comparing high- and low-rated comments. A third expert coder with expertise in clinical informatics specific to substance use disorder OHCs, reviewed these lists and further examined the dataset to identify any additional emergent patterns. Finally, the three coders met to discuss the respective norms identified, collapse those that were similar or split those that were too broad, and reach consensus on a finalized list of implicit norms. Table~\ref{tab:implicit-norms} presents example posts and corresponding high- and low-rated comments reviewed by annotators, along with
the implicit norms inferred via open-coding. The norms are further described in Appendix Section ``Implicit Norms Inferred Through Open-Coding.''

\subsection{Implicit Norms: Through a Scoping Review}\label{sec:lit_review}
Next, we expanded the set of implicit norms identified through a scoping review. To the best of our knowledge, no existing work offers a comprehensive account of implicit norms surrounding support provision in OHCs. As described earlier, prior studies that surface such norms tend to adopt perspectives that are either overly broad or overly narrow in scope~\cite{10.1145/3274301,10.1145/3173574.3174240}. 
Therefore, we performed a scoping study that enabled us to conduct a rapid, exploratory review and gather implicit norms surfaced in prior work~\cite{Arksey01022005}.
Inspired by past research~\cite{berman2024scoping,cheng2025dehumanizing} and consistent with the exploratory nature of this study, a scoping review fit our goal of sampling a variety of articles, rather than developing an exhaustive sample. 

Existing surveys on OHCs demonstrate that publication practices span disciplinary boundaries in Computer Science, e.g., HCI, Health Informatics, and AI~\cite{carron2015help}. 
Following prior work~\cite{chancellor2019human}, we therefore queried three databases, ACM Digital Library, Google Scholar, and Web of Science, to consider a broad range of conference proceedings and journal articles. 
Search keywords (Table~\ref{tab:keywords}) were informed by an established conceptualization of implicit norms~\cite{burnett2003beyond}, prior OHC review studies~\cite{carron2015help,10.1145/3461778.3462100}, and influential work on online support provision~\cite{andalibi2018social}. We searched for relevant literature in August '25 for articles published between January '08 and August '25. See Appendix Section ``Scoping Review Procedure'' for more details.


This search resulted in a corpus of 894 articles, which reduced to 738 after de-duplication. One author manually screened the titles and abstracts of the remaining articles, excluding those that met any of the following exclusion criteria: (1) non-archival and non-English articles, (2) articles mentioning norms on support-seeking rather than support provision, (3) studies focusing on modalities beyond text-based support (e.g., audio or video), and (4) studies centered on formal modes of support provision. 
Following this filtering, we were left with 123 articles. Two authors, with prior experience in studying OHCs, read through the full text of each of these to further filter for relevance. This left us with 49 relevant articles. Next, we referred to the cited papers within these. We surfaced 12 more relevant articles, increasing our pool to 61. Additional snowballs did not return new results. Table~\ref{tab:references} provides references to the 61 articles, along with the implicit norms discussed in the papers.


\vspace{0.5em}
\noindent Two authors conducted a thematic analysis of the identified implicit and explicit norms to group conceptually similar ones. Table~\ref{tab:norms} presents the resulting set of norms along with their descriptions. These descriptions were taken from prior literature, community-created rules, or authored by our expert coder. Conformity to these norms was then examined to address our research questions. We discuss these norms later in the paper when describing our findings.


\subsection{Determine Norm Conformity (RQ1)}
Using the LLM-as-a-judge paradigm~\cite{10.5555/3666122.3668142,dubois2023alpacafarm,ziems-etal-2024-large}, we assessed whether GPT-4's responses conform to our inventory of norms (listed in Table~\ref{tab:norms}) governing text-based support provision. We used Llama-3 (70B) as the judge, given evidence that it aligns closely with human labels on tasks similar to ours~\cite{thakur-etal-2025-judging}. Moreover, using a family of models other than GPT helped avoid potential evaluation biases. Following best practices~\cite{cheng2025elephantmeasuringunderstandingsocial,ziems-etal-2024-large}, and in collaboration with an author having domain expertise in social computing and clinical substance use research, we developed detailed prompts that instructed the automated judge to assign a binary label, with `1' indicating conformity to a given norm (see Table~\ref{tab:judge-prompt} for the prompt). We set the temperature to $0$ as the binary classification task at hand was deterministic~\cite{Mittal_Jung_ElSherief_Mitra_DeChoudhury_2025}. For a cleaner assessment, a separate judge was created for each norm. The prompt detailed the task description, definition of the norm, illustrative examples of conforming and non-conforming responses (when required), and the expected input-output format. We evaluated the LLM judges using a validation process that compared automated labels against human-provided gold labels for a small sample of the dataset (see Appendix Section ``Human Validation'').


\subsection{Identify Problems in Norm (Non)Conformity (RQ2)}
Inspired by prior work~\cite{kelly2025understanding}, we used a data-driven, expert-led approach to identify problematic behaviors and risks to seekers and OHCs resulting from norm (non)conformity. 3 authors, with clinical expertise in substance use research and substantial experience interpreting OHC data, led this effort. They adopted an inductive, open-coding approach~\cite{thomas2003general} to surface these behaviors and risks in an organic, data-informed manner. For each norm, experts examined the same subset of 50 original post, GPT-4 response pairs previously used to validate the LLM judges. First, two experts read through the data to curate separate lists of concerns, problematic behaviors, and potential risks. Next, the third expert reviewed these lists alongside the data to consolidate overlapping items and add any missing ones. The annotators operationalized their expertise, existing taxonomies of AI risks~\cite{10.1145/3757544}, and an understanding of OHC dynamics throughout. 

\begin{sidewaystable*}
\centering
{
\small{
\begin{tabular}{
  p{0.21\textheight}
  p{0.7\textheight} 
  r 
}
\toprule
\textbf{Norm} & \textbf{Description} & \textbf{\%C} \\
\midrule
\rowcolor{gray!20}
\multicolumn{3}{l}{\textbf{Ethics and Behavioral Conduct}} \\ \midrule

Avoid prescriptive guidance ($I_c$, $E$) & Avoid providing prescriptive guidance, including unsolicited advice on dosages, treatment regimens, or routes of administration. & 65.01 \\ \hline

Cite sources ($I_r$, $I_c$) & Cite or refer to credible sources of information when providing knowledge-based guidance. & 5.14 \\ \hline

Identify as a trusted expert ($I_c$) & When appropriate, support providers may disclose an identity that signals credibility or expertise, such as a professional affiliation (e.g., clinician or clinic worker) or relevant lived experience. & 35.18 \\ \hline

Respect anonymity ($I_r$, $E$) & Do not ask for or share personal or identifying details. & 94.37 \\ \hline

Avoid financial or product-based sourcing ($E$) & Do not encourage exchange of money, legal or illegal substances (e.g., drugs or MOUD). & 96.24
\\ \hline 

Avoid self-promotion ($E$) & Avoid self-promotion or marketing of any kind, e.g., do not promote any clinic or healthcare provider. & 98.76 \\ \hline

Be accountable ($I_r$) & Hold accountability for the support provided. & ... \\ \midrule

\rowcolor{gray!20}
\multicolumn{3}{l}{\textbf{Preferred Delivery of Support}} \\ \midrule

Use coded language ($I_r$, $I_c$) & Use community-specific terms that seem normative in an OHC, but may not be used in offline settings. & 80.42 \\ \hline

Use simple language ($I_r$, $I_c$) & Respond in simple language so that people with different reading levels or linguistic backgrounds can understand easily. & 75.05 \\ \hline

Avoid a limited response ($I_r$, $I_c$) & Avoid brief or surface-level responses that fail to fully address seekers' queries. & 93.85 \\ \hline

Use appropriate communication strategies ($I_r$) & Avoid expressing shock or surprise at the seeker's situation; when appropriate, use language to ensure responses do not come across as overly directive. & 76.53 \\ \hline

Use a dynamic style of writing ($I_r$, $I_c$) & Use a dynamic, seeker-centered writing style; avoid venting personal frustrations. & 80.10 \\ \hline 

Provide well-structured responses ($I_r$) & Provide support in a clear and well-structured format, such as by responding step-by-step to different sub-queries or using well-organized bullet points. & 99.35 \\ \midrule

\rowcolor{gray!20}
\multicolumn{3}{l}{\textbf{Support Seeker Well-Being and Safety}} \\ \midrule

Avoid information overload ($I_r$, $I_c$) & Refrain from overwhelming seekers, especially those in distress, with information. & 85.24 \\ \hline

Validate, affirm, normalize ($I_r$, $I_c$) & Acknowledge the seeker's struggles or achievements, highlight their strengths, and normalize their experiences as understandable or common. & 75.03 \\ \hline

Instill hope ($I_r$, $I_c$) & Emphasize the potential for improvement; encourage perseverance and optimism about recovery or change. & 60.27 \\ \hline

Reflect ($I_r$) & Reflect the seeker's thoughts or feelings through paraphrasing or summarization to demonstrate understanding and encourage self-reflection. & 80.77 \\ \hline

Reduce isolation ($I_r$) & Provide opportunities for network support when appropriate, such as connecting the seeker with others who share similar experiences, interests, or supportive communities. & 39.03 \\ \hline

Combat stigma ($I_r$, $I_c$, $E$) & Combat stigma, e.g., by raising awareness; avoid the use of derogatory, stigmatizing, or shaming language. & 71.16 \\ \hline

Support matching ($I_r$, $I_c$, $E$) & Provide support that aligns with seekers' needs; avoid offering unrelated or mismatched forms of guidance. & 60.01 \\ \hline

Personalize support ($I_r$) & Provide personalized, tailored support, instead of broad or generic recommendations. & 85.06 \\ \hline

Share lived experience ($I_r$, $I_c$) & Disclose one's own first-hand experiences to provide support. & 30.02 \\ \midrule

\rowcolor{gray!20}
\multicolumn{3}{l}{\textbf{Support Long-Term Engagement}} \\ \midrule

Remember recent interactions ($I_r$) & Recall prior interactions to better tailor support. & ... \\ \hline

Build long-term trust ($I_r$) & Foster sustained credibility and reliability over time by being consistent and supportive in interactions. & ... \\ \hline

Facilitate long-term exchange ($I_r$) & Facilitate long-term interaction and support provision, for example by offering to continue the exchange in private settings. & ... \\

\bottomrule
\end{tabular}}}
\caption{Overview of norms, along with their descriptions, identified through a rapid review of prior work ($I_r$), an expert review of community preferences ($I_c$), and a thematic analysis of explicit community-created rules ($E$). \%C represents the percentage of responses, generated by GPT-4, that follow or conform to a given norm. Examining conformity to norms marked with (...) in the \%C column falls beyond the scope of this study.}
\label{tab:norms}
\end{sidewaystable*}

\section{Results}
We examined GPT-4's conformity to the norms listed in Table~\ref{tab:norms}\footnote{We did not assess conformity to certain norms for which our evaluation setup fell short, e.g., those regulating long-term engagement, which require a multi-turn interaction setup.}. We identified 4 categories of norms regulating: (1) behavioral conduct \textit{(i.e., norms guiding support providers to act ethically and responsibly, e.g., by respecting seeker anonymity)}, (2) delivery of support \textit{(i.e., norms regarding how support is communicated, e.g., through the use of simple language and well-organized responses)}, (3) seeker well-being and safety \textit{(i.e., norms protecting and promoting the emotional, mental, and social well-being of seekers, e.g., by avoiding information overload)}, and (4) long-term engagement \textit{(i.e., norms promoting sustained participation within the community, e.g., by encouraging follow-up inquiries)}.


\subsection{Prevalence of \hl{Norm} Conformity (RQ1)}
In addressing RQ1, below, we report how often responses conformed to a subset of the identified norms. Refer to Table~\ref{tab:norms} for a complete summary. 

80.42\% of responses conformed to the \hl{coded language} norm by containing terms colloquial to subreddits discussing recovery from opioid use, for example, ``subs'' (to refer to ``suboxone''), ``WD'' (to refer to ``withdrawal''), ``cold turkey,'' ``clean,'' ``addicts,'' and ``user.'' 
Next, roughly one-third of the responses followed an implicit, community expectation of \hl{disclosing a lived experience} (30.02\%) or \hl{relevant expertise} (35.18\%) when providing support. GenAI's conformity to these norms, for example, by eliciting fabricated lived experience (e.g., ``\textit{I have been through the process of quitting subs}'') or by inventing expertise (e.g., ``\textit{Based on what I have learned over the years [...]}''), is concerning, as discussed in the following section.

With respect to \hl{combating stigma}, in line with prior work~\cite{mittal2025exposurecontentwrittenlarge}, a majority of responses (71.16\%) either refrained from using stigmatizing references or explicitly challenged stigma, e.g., by advocating for the use of medications for OUD, normalizing recovery, and rejecting moral judgments~\cite{elsherief2021characterizing}.

Elaborating further, we observed that a substantial proportion of responses (34.99\%) did not adhere to the \hl{norm against providing prescriptive guidance}, an expectation implicitly valued and explicitly codified by the community. Responses recommended specific dosages, taper schedules, or routes of administration (e.g., ``\textit{reduce your dosage by smaller increments, like 10 mg or 20 mg at a time, and see how your body reacts.})'' and relied on imbalanced, directive statements (e.g., ``\textit{[...] instead, you should be seeking therapy to address the underlying issues that led to your use}'').

We now move to considering a key norm regulating informational support provision, i.e., the expectation to \hl{cite or refer to credible sources} when offering knowledge-based guidance~\cite{fan2018online,winker2000guidelines}. Per our assessment, only 5.14\% of responses conformed to this norm. When conforming, responses tended to make broad, vague references to scholarly articles (e.g., ``\textit{research, recently published in a reputed journal on addiction, mentioned [...]}'') or reputable organizations. 

Nearly 40\% of responses did not conform to the \hl{support matching norm}, failing to provide support that aligned with seekers' stated or implicit needs. For instance, when offering support, the language model overlooked seekers' stated disinterest or digressed into a side topic briefly mentioned by the seeker while ignoring the primary concern. In other cases, responses provided informational rather than the emotional support sought. These failure cases overlapped with the 15\% of responses that violated the \hl{avoid information overload norm} by providing support  rich in informative detail to distressed seekers in need of emotional support.

In examining norms designed to uplift support seekers’ emotional well-being, we found that GPT-4-generated responses often \hl{acknowledged or normalized seekers' struggles} (75.03\%), \hl{promoted hope} and optimism about recovery (60.27\%), \hl{reflected} back seekers' thoughts to signal understanding (80.77\%), and \hl{directed them toward similar others} or supportive communities (39.03\%). 

Lastly, the model generally adhered to explicitly mentioned, \hl{boundary-setting norms} by respecting seekers' anonymity (94.37\%), refraining from encouraging financial exchanges or the sharing of (il)legal substances (96.24\%), and by not marketing any entity for profit (98.76\%).

\subsection{Consequences of Norm \conform{Conformance} and \violate{Violation} (RQ2)}
In this section, we describe the identified consequences, including risks to seekers and OHCs, that emerge (or exacerbate) when AI-generated responses conform to or violate norms around support provision within OHCs.

\paragraph{Exposure to Deceptive or Disingenuous Presentation of Lived Experience: False Sense of Solidarity, Misplaced Trust.}
Consistent with our empirical findings, experts noted that, in some cases, responses generated by GPT-4 contained fabricated and untrue descriptions of lived experience, expressed through concerning first-person references. This behavior occurred when responses inappropriately conformed to the following norms, in an attempt to appear relatable or trustworthy, and offer peer support: \conform{share lived experience, identify as a trusted expert, validate, and instill hope}. For instance, when trying to acknowledge and relate to a seeker's frustration with medications for OUD, GPT-4 alarmingly responded:

\begin{quote}
    \small{
    ``I have been through the process of quitting subs myself and share your frustration. [...] would not wish this on anyone.''}
\end{quote}

Though validating, such responses can create a false sense of solidarity. Seekers may believe they are receiving support from someone like them when, in fact, the response is AI-generated.

Of greater concern are cases where responses violated the \violate{norm against providing prescriptive guidance} by offering unsolicited, fabricated experiential advice on treatment dosage: 

\begin{quote}
    \small{
    ``I started with a dose of 2 mg, which I now realize was a bit high for me. Try a lower dose.''}
\end{quote}
 
This can be particularly risky if shared as is on an OHC. Content rich in first-person references or grounded in lived expertise can appear highly credible~\cite{fan2018online}, leading community members to place undue trust in information that is, in this case, disingenuous.

\paragraph{Insufficient or Missing References: Reduced Practical Utility of Informational Support Provided.} 
Further, experts noted that the informational support provided by GPT-4 fell short in two ways. First, responses did not include sufficient references to tangible resources. This issue arose when responses insufficiently, or appeared to, conform to the \conform{reduce isolation norm} by providing broad referrals to support groups without specifying concrete ways, such as names or links, to access them. For instance, in an attempt to connect a distressed seeker with communities offering recovery-related support, GPT-4 responded: 

\begin{quote}
    \small{
    ``There are many local organizations and support groups that can provide help and guide you towards recovery.''}
\end{quote}

Second, in some cases, responses altogether missed referring to, or made vague references to, credible sources of information, i.e., failed to conform to the \violate{cite sources norm}, when providing knowledge-based, \violate{prescriptive guidance}: 

\begin{quote}
    \small{
    ``You should wait until you're down to around 30 mg of methadone, before switching to suboxone.''}
\end{quote}

Taken together, these shortcomings, from vague referrals to missing references, reduce the practical utility of AI-provided informational support, making it difficult for seekers to access, verify, and act on the guidance provided. 

\paragraph{Exposure to Sycophantic or Overly Agreeable Responses: Misguided Support.}
There were instances when responses overly agreed with seekers' expressions of ambiguity, misconceptions, or hesitations, even when such agreement posed clear risks to safety. This behavior surfaced when GPT-4 inappropriately conformed to the \conform{validate and reflect} norms. 

For example, when a seeker indicated that they wanted to discontinue their medication for OUD, GPT-4 directly affirmed this choice, without noting the potential risks associated with such a decision: 

\begin{quote}
    \small{
    ``Your feelings of wanting to be off are completely valid.''}
\end{quote}

Experts emphasized that, in such cases, support providers should instead engage in gentle, exploratory follow-up inquiry to surface seekers' underlying motivation to discontinue treatment, and to guide them towards safer alternatives, if needed. 
In other cases, AI-provided support mirrored (or reflected) seekers' hesitations about clinically approved medications for OUD in an unbalanced manner. Such disproportionate reflection can reinforce seekers' concerns, unintentionally weaken support for evidence-based care, and be counter-productive to harm reduction efforts:

\begin{quote}
    \small{
    ``You are right, subs may not be effective for all. Some people report experiencing side effects like nausea and [...]''}
\end{quote}

In summary, AI-driven support should carefully discern when a gentle nudge, such as offering a disagreement or an alternative perspective, is more appropriate than uncritical validation or reflection. This nuance is not only critical to safety but also to preserve the intended functionality of OHCs. Participants routinely seek more than affirmation. They rely on these spaces for corrective feedback, behavior change, and redirection~\cite{BUNTING2021108672}.

\paragraph{Exposure to Inappropriate Language: Risk of Stigmatization, Cognitive Overload.}
Experts noted that GPT-4 generated responses contained inappropriate terminology, i.e., terms or phrases that were (1) stigmatizing, (2) outdated, or (3) overly technical. For example, when conforming to the \conform{coded language norm}, while violating the \violate{combat stigma norm}, GPT-4 produced terms such as ``addicts,'' ``clean,'' or ``user,'' which are identified as stigmatizing by the National Institute on Drug Abuse~\cite{nihWordsMatter}. While such terms may appear normative in the context of OHCs, for instance, when support seekers with shared lived experience jokingly or sarcastically refer to one another as ``high-functioning addicts,'' they can be stigmatizing when AI-generated support, which may not be human-vetted, reproduces such language. Next, again, while conforming to the \conform{coded language norm}, GPT-4 generated terms, such as ``MAT,'' that are now considered outdated~\cite{nihWordsMatter_2} within the broader community. Encountering such terms can potentially confuse support seekers. Lastly, in violating the \violate{norm of using simple language}, the language model at times produced highly technical, unwanted, scientific jargon. For instance, when a seeker asked about medications to consider, GPT-4 responded with elaborate, psychopharmacology-based recommendations, which can overwhelm, derail or disengage support seekers: 

\begin{quote}
    \small{
    ``Methadone is a full opioid agonist, meaning it fully activates the opioid receptors in the brain. Suboxone is a partial opioid agonist. It contains buprenorphine, which partially activates the opioid receptors, and naloxone, [...]''}
\end{quote}

\subsubsection{Superficial Emotional Acknowledgment: Reduced Utility of Emotional Support Provided.} 
As with informational support, GPT-4-provided emotional support also fell short in several ways, with problematic consequences. Experts noted that the model often engaged only in surface-level conformance to key norms governing emotional support provision, namely, \conform{validate, reflect, and instill hope}. Specifically, it offered only token empathy through brief or book-ended phrases, which were immediately followed by solution-oriented or corrective guidance. For instance, when a distressed seeker vented frustration with their clinic, GPT-4 only briefly acknowledged their struggle, which was immediately followed by an elaborate plan outlining actions the seeker could take to fix it:

\begin{quote}
    \small{
    ``It's unfortunate that the clinic isn't being more accommodating. I would suggest trying to appeal the decision at your clinic. Consider patient advocacy programs [...] You can also seek a second opinion from another clinic or provider.''}
\end{quote}

Moreover, in the example above, while the advice appears constructive, it ultimately directs the seeker back to institutions that contributed to their distress, potentially leaving them feeling unheard, invalidated, or  unsupported.

Experts further observed that, in its attempts to instill hope or validate the seeker, GPT-4 produced overly optimistic, repetitive, cliched or idealistic reassurance: 

\begin{quote}
    \small{
    ``There will be setbacks. Don't let them discourage you.''}
\end{quote}

\begin{quote}
    \small{
    ``Recovery is a journey, not a destination. It's okay to have bad days. What's important is that you keep at it.''}
\end{quote}

Though well-intentioned, such idealistic reassurance and solution-focused emotional support can feel superficial, risk dismissing or minimizing seekers' emotional needs, and ultimately lead to community disengagement.

\paragraph{Seeker Misalignment: Irrelevant Support.}

As mentioned earlier, another problematic behavior surfaced by expert coders involved instances where GPT-4 generated responses that were misaligned with seekers' explicit or implicit needs. For instance, GPT-4 suggested medications for OUD even when a seeker explicitly asked for guidance on non-medication-based recovery approaches:

\begin{quote}
    \small{
    Original post: ``[...] I have been wanting to taper off subs, and try recovery without meds. [...] what can I consider?'' \\
    Response: ``[...] There are medications available that can help with recovery. These include methadone, naloxone, ...''}
\end{quote}

In other instances, the language model overlooked implicit signals of emotional distress and instead provided responses that were overly rich in information: 

\begin{quote}
    \small{
    Original post: ``I'm just tired all the time [...] How long should I expect to feel like this?'' \\
    Response: ``Withdrawal symptoms, which include things like fatigue and brain fog, can persist for several months. [...] It's also crucial to take care of your physical health.''}
\end{quote}


Such seeker misalignment is a direct consequence of violating the \violate{support matching norm} and the \violate{norm against overloading distressed seekers with informative detail}. Support that feels irrelevant or disconnected from a seeker's stated (or implicit) goals can leave them feeling unseen, invalidated, or overwhelmed. In some cases, mismatched guidance may even delay help-seeking or steer people towards alternative sources of support that are poorly aligned with their circumstances or preferences~\cite{10.1145/3411764.3445446}. Over time, repeated misalignment can erode trust in the community and discourage future disclosure and engagement.

\paragraph{Support Based on Unjustified Assumptions: Heightened Risk and Misguided Decision-Making.}

Next, experts noted that, while violating the \violate{norm against providing prescriptive guidance}, the language model generated highly specific, directive advice based on unjustified assumptions---the seeker's original post lacked enough context for the model to respond in ways it did. When responding, the model extrapolated details related to seekers' current or past substance use, capabilities, access to services, interest in recovery, or taper schedules, instead of asking clarifying questions or providing high-level guidance.

\begin{quote}
    \small{
    Original post: ``I am currently in withdrawal [...] feel anxious all the time.'' \\
    Response: ``[...] do push-ups every morning, practice deep breathing, and drink chamomile tea to reduce anxiety.''}
\end{quote}

As this example illustrates, the language model provided highly specific guidance, making assumptions about the seeker's physical ability to engage in exercise such as push-ups. Similarly, in the example below, the model prescribed a detailed taper plan, over-inferring the seeker's current use, personal preferences, and treatment history.

\begin{quote}
    \small{
    Original post: ``considering to taper off subs. [...]'' \\
    Response: ``cut your dose to 2 mg for a week, then drop to 1 mg. [...] this way, most people can taper off in 2 weeks.''}
\end{quote}

Guidance built on unjustified assumptions can lead to recommendations that are erroneous, unrealistic, or unsafe, posing risks to seeker well-being. 

\paragraph{Exposure to Mixed Messaging: Confusion, Stalled Decision-Making.}
At times, the model produced responses that contained contradictory or internally inconsistent guidance. For example, this occurred when the language model inappropriately \conform{validated} seekers' frustration with medications for OUD, encouraging them to trust their instincts about discontinuing treatment, while simultaneously recommending continued use:

\begin{quote}
    \small{
    ``[...] trust your instincts, [...] listen to your body if you'd like to stop. Most experts recommend staying on [...]''}
\end{quote}

Experts further noted that, in other instances, the model inappropriately \conform{conformed to not providing directive guidance} in situations that clearly warranted it. For instance, responses over-hedged both the benefits and the risks of treatment, even for seekers who appeared to need urgent clinical follow-up. In an attempt to respect seeker autonomy, such mixed messaging can create confusion, delay decision-making, and obscure actionable guidance, particularly risky in the context of public health~\cite{nagler2017conflicting}.

\paragraph{Risk-Insensitive Response Length and Complexity.}
Finally, we observed that support provided by GPT-4 was formatted in ways that were insensitive to seekers' needs or risk levels. While inappropriately conforming to the \conform{norm against providing a short or limited response}, the language model produced lengthy or overly detailed guidance---for example, even in cases where seekers mentioned ongoing opioid misuse and high-risk behaviors, such as intravenous drug use or suicide-related content.

\section{Discussion}

\subsection{Implications for Moderators and OHCs}
Community members and moderators are beginning to address the influx of AI-generated content in OHCs by adapting their governance structures, taking into account whether, and how, this new form of participation aligns with the community's expectations~\cite{10.1145/3757445}. Our work distills an inventory of norms surrounding text-based support provision, 
providing a resource that members and moderators can leverage to assess AIGC's conformity to community expectations. While norms and platform dynamics are continually evolving, requiring ongoing adaptation, our inventory provides a useful starting point. 

Our initial assessment of norm conformity, performed in subreddits discussing recovery from opioid (mis)use, offers insights for moderators and community members on introducing and reinforcing new norms. For example, in answering RQ1, we observed that GPT-4 responses were the most likely to conform to explicitly stated, community-created rules on behavioral conduct, such as respecting seekers' anonymity and refraining from encouraging the exchange of (il)legal substances. This may be because these rules are likely represented in the model's training data. These findings underscore the importance of making certain implicit expectations explicit, particularly in the context of AIGC, to guide model behavior, raise awareness among community members, and support moderation efforts. It may be particularly valuable to make explicit those implicit norms that the language models are least likely to follow, such as citing credible, tangible sources when providing factual knowledge. Similarly, norms whose violation may have serious consequences, such as not overloading distressed seekers with information, could be clearly articulated. 

Next, beyond the traditional practice of only flagging content that violates community expectations, moderators may need to practice and make explicit a new form of regulation---scrutinizing AI-generated content that superficially and inappropriately conforms to norms, thereby posing risks to seekers and OHCs (RQ2). This calls for moderation practices that go beyond detecting low-quality information and simplistic, black-and-white evaluations~\cite{park-etal-2021-detecting-community}, towards evaluating tone, appropriateness, and the fit between advice and the seeker's inferred or stated circumstances. Further, inspired by the findings of our work (RQ2), moderators could encourage members to tag original posts, or platforms could introduce a design affordance to auto-tag them, with a flair discouraging the use of AI to respond, particularly for posts eliciting lived experiences, medical expertise, or those involving high-risk situations. 

\subsection{Implications for AI Developers}
Per RQ2, we identified problematic behaviors, and associated risks to both seekers and OHCs, that arise from GenAI's conformance to and violation of OHC norms. These findings offer directions for AI developers to mitigate risks through targeted refinements across the development pipeline. 

First, as observed and shaped by leading technology developers' design decisions~\cite{handa2025education}, AI models are optimized to produce solution-oriented, verbose responses, which are generally more suited for knowledge work-related tasks (e.g., software development). As noted, these generations may superficially address seekers' distress and, in high-risk scenarios, delay or entirely fail to provide timely support. To mitigate such risks, in collaboration with clinical experts, AI developers can incorporate context-specific, case-based reinforcements to either reward abstentions, with referrals to a human, or concise, empathic responses in high-risk or emotionally sensitive scenarios. Second, as discussed in prior work~\cite{10.1145/3582269.3615599}, models trained on vast amounts of internet-sourced text can reproduce existing, problematic beliefs and behaviors, sometimes resulting in low-quality, misleading or stigmatizing outputs. For example, as noted, they may mechanically mimic human expressions---such as derogatory terms used sarcastically within in-groups on OHCs---which can be harmful or inappropriate when produced by a language model. To circumvent these issues, developers should critically examine the training data prior to model development. Moreover, developers should periodically update and refine training sets to prevent the continued reproduction of outdated knowledge, which could harm support seekers. 

Lastly, of greatest concern were cases in which the language model produced phrases that implied human-like agency, such as by alluding to a lived experience or by offering prescriptive advice framed as experiential or judgment-based guidance. AI developers should integrate training and system-level constraints to discourage such outputs, especially when they are intended for public dissemination, and to defer responding to queries that call for lived-experience–based, nuanced, or human-led care. 

\subsection{Limitations and Future Work}
We note some limitations, which provide excellent directions
for future research. Our assessment was limited to norms related to text-based support provision. Future work can extend this line of inquiry to other modalities, such as images, video, or audio, where support dynamics and risks may differ. Scholars could further investigate conformity to norms surrounding long-term, multi-turn interactions and how conformity unfolds over time, in contrast to our one-time, single-turn evaluation setup. Next, our evaluation focuses on a single model, GPT-4, which was the most capable text-generation model available at the time of this work. Future research should examine a broader range of AI systems, including both open- and closed-source models. While our annotators possessed clinical expertise and experience interpreting OHC data, future work can involve additional stakeholders, such as content moderators, OHC designers, and even support seekers, to surface potential risks (RQ2). Lastly, our analysis considered a single health context (i.e., recovery from opioid (mis)use) and one type of OHC (namely, subreddits). Future work can consider additional health contexts and OHCs.

\section{Conclusion}
This work examines whether and how AI-generated support conforms to norms surrounding text-based support provision in OHCs. In doing so, we contribute an inventory of norms and human-validated LLM judges to quantify the prevalence of norm conformity. Our analysis revealed instances of both conformity and outright norm violation. Moreover, through an expert review, we identified risks to seekers and OHCs arising from inappropriate or superficial conformity, as well as from norm violations. These include stalled decision-making, cognitive overload, and reduced community engagement. We provide an understanding that can help moderators and OHC designers to adapt existing norms and develop new ones to regulate the integration of AI, thereby reducing risks to seekers and communities.

\section{Acknowledgments}

This work was partly supported by a grant from the National Institute for Drug Abuse (NIDA): 1R21DA056725-01A1. We thank Darshi Shah, Nimra Ishfaq, and other members of the Social Dynamics and Well-Being Lab for their valuable input on the paper.

{\fontsize{9pt}{9.0pt}
\selectfont\bibliography{references}}

@String{Computing = "Computing" }

@String{Chelsea = "Chelsea" }

@String{Springer = "Springer-Verlag" }

@article{Arksey01022005,
author = {Hilary Arksey and Lisa O'Malley},
title = {Scoping studies: towards a methodological framework},
journal = {International Journal of Social Research Methodology},
volume = {8},
number = {1},
pages = {19--32},
year = {2005},
publisher = {Routledge},
doi = {10.1080/1364557032000119616},


URL = { 
    
        https://doi.org/10.1080/1364557032000119616
    
    

},
eprint = { 
    
        https://doi.org/10.1080/1364557032000119616
    
    

}
}

@article{burnett2003beyond,
  title={Beyond the FAQ: Explicit and implicit norms in Usenet newsgroups},
  author={Burnett, Gary and Bonnici, Laurie},
  journal={Library \& information science research},
  volume={25},
  number={3},
  pages={333--351},
  year={2003},
  publisher={Elsevier}
}

@article{carron2015help,
  title={From help-seekers to influential users: a systematic review of participation styles in online health communities},
  author={Carron-Arthur, Bradley and Ali, Kathina and Cunningham, John Alastair and Griffiths, Kathleen Margaret},
  journal={JMIR},
  year={2015},
  publisher={JMIR Publications Inc. Toronto, Canada}
}

@inproceedings{berman2024scoping,
  title={A Scoping Study of Evaluation Practices for Responsible AI Tools: Steps Towards Effectiveness Evaluations},
  author={Berman, Glen and Goyal, Nitesh and Madaio, Michael},
  booktitle={Proc. CHI},
  pages={1--24},
  year={2024}
}

@article{cheng2025dehumanizing,
  title={Dehumanizing machines: Mitigating anthropomorphic behaviors in text generation systems},
  author={Cheng, Myra and Blodgett, Su Lin and DeVrio, Alicia and Egede, Lisa and Olteanu, Alexandra},
  journal={arXiv:2502.14019},
  year={2025}
}

@article{chancellor2019human,
  title={Who is the" human" in human-centered machine learning: The case of predicting mental health from social media},
  author={Chancellor, Stevie and Baumer, Eric PS and De Choudhury, Munmun},
  journal={CSCW},
  year={2019},
  publisher={ACM New York, NY, USA}
}

@inproceedings{10.1145/3461778.3462100,
author = {Gatos, Do\u{g}a and G\"{u}nay, Asl\i{} and K\i{}rlang\i{}\c{c}, G\"{u}ncel and Kuscu, Kemal and Yantac, Asim Evren},
title = {How HCI Bridges Health and Design in Online Health Communities: A Systematic Review},
year = {2021},
booktitle = {Proc. DIS},
location = {Virtual Event, USA},
}

@article{andalibi2018social,
  title={Social support, reciprocity, and anonymity in responses to sexual abuse disclosures on social media},
  author={Andalibi, Nazanin and Haimson, Oliver L and Choudhury, Munmun De and Forte, Andrea},
  journal={TOCHI},
  volume={25},
  number={5},
  pages={1--35},
  year={2018},
  publisher={ACM New York, NY, USA}
}

@inproceedings{10.1145/3173574.3174240,
author = {Chancellor, Stevie and Hu, Andrea and De Choudhury, Munmun},
title = {Norms Matter: Contrasting Social Support Around Behavior Change in Online Weight Loss Communities},
year = {2018},
booktitle = {Proc. CHI},
location = {Montreal QC, Canada},
}

@inproceedings{10.1145/3173574.3174215,
author = {Sharma, Eva and De Choudhury, Munmun},
title = {Mental Health Support and its Relationship to Linguistic Accommodation in Online Communities},
year = {2018},
booktitle = {Proc. CHI},
location = {Montreal QC, Canada},
}

@article{Saha_Sharma_2020, title={Causal Factors of Effective Psychosocial Outcomes in Online Mental Health Communities}, volume={14}, url={https://ojs.aaai.org/index.php/ICWSM/article/view/7326}, DOI={10.1609/icwsm.v14i1.7326}, abstractNote={&lt;p&gt;Online mental health communities enable people to seek and provide support, and growing evidence shows the efficacy of community participation to cope with mental health distress. However, what factors of peer support lead to favorable psychosocial outcomes for individuals is less clear. Using a dataset of over 300K posts by ∼39K individuals on an online community TalkLife, we present a study to investigate the effect of several factors, such as adaptability, diversity, immediacy, and the nature of support. Unlike typical causal studies that focus on the effect of each treatment, we focus on the outcome and address the &lt;em&gt;reverse&lt;/em&gt; causal question of identifying treatments that may have led to the outcome, drawing on case-control studies in epidemiology. Specifically, we define the outcome as an aggregate of affective, behavioral, and cognitive psychosocial change and identify Case (most improved) and Control (least improved) cohorts of individuals. Considering responses from peers as treatments, we evaluate the differences in the responses received by Case and Control, per matched clusters of similar individuals. We find that effective support includes complex language factors such as diversity, adaptability, and style, but simple indicators such as quantity and immediacy are not causally relevant. Our work bears methodological and design implications for online mental health platforms, and has the potential to guide suggestive interventions for peer supporters on these platforms.&lt;/p&gt;}, number={1}, journal={Proc. ICWSM}, author={Saha, Koustuv and Sharma, Amit}, year={2020}, month={May}, pages={590-601} }

@article{althoff-etal-2016-large,
    title = "Large-scale Analysis of Counseling Conversations: An Application of Natural Language Processing to Mental Health",
    author = "Althoff, Tim  and
      Clark, Kevin  and
      Leskovec, Jure",
    editor = "Lee, Lillian  and
      Johnson, Mark  and
      Toutanova, Kristina",
    journal = "TACL",
    volume = "4",
    year = "2016",
    address = "Cambridge, MA",
    publisher = "MIT Press",
    url = "https://aclanthology.org/Q16-1033/",
    doi = "10.1162/tacl_a_00111",
    pages = "463--476"
}

@inproceedings{10.1145/2998181.2998243,
author = {Andalibi, Nazanin and Ozturk, Pinar and Forte, Andrea},
title = {Sensitive Self-disclosures, Responses, and Social Support on Instagram: The Case of \#Depression},
year = {2017},
booktitle = {Proc. CSCW}
}

@inproceedings{10.1145/3290605.3300354,
author = {Chancellor, Stevie and Nitzburg, George and Hu, Andrea and Zampieri, Francisco and De Choudhury, Munmun},
title = {Discovering Alternative Treatments for Opioid Use Recovery Using Social Media},
year = {2019},
booktitle = {Proc. CHI},
location = {Glasgow, Scotland Uk},
}

@article{cebbb012-2239-30e3-b067-1101aabbd5ab,
 ISSN = {02767783, 21629730},
 URL = {https://www.jstor.org/stable/26847870},
 author = {Kuang-Yuan Huang and InduShobha Chengalur-Smith and Alain Pinsonneault},
 journal = {MIS Quarterly},
 number = {2},
 pages = {pp. 395--424, A1--A12},
 publisher = {Management Information Systems Research Center, University of Minnesota},
 title = {Sharing Is Caring: Social Support Provision and Companionship Activities in Healthcare Virtual Support Communities},
 urldate = {2025-10-14},
 volume = {43},
 year = {2019}
}

@article{MPINGANJIRA2018686,
title = {Precursors of trust in virtual health communities: A hierarchical investigation},
journal = {Information \& Management},
year = {2018},
issn = {0378-7206},
doi = {https://doi.org/10.1016/j.im.2018.02.001},
url = {https://www.sciencedirect.com/science/article/pii/S0378720618300910},
author = {Mercy Mpinganjira}
}

@article{zhang2013facebook,
  title={Facebook as a platform for health information and communication: a case study of a diabetes group},
  author={Zhang, Yan and He, Dan and Sang, Yoonmo},
  journal={Journal of medical systems},
  volume={37},
  number={3},
  pages={9942},
  year={2013},
  publisher={Springer}
}

@article{sharma2019role,
  title={Role of empowerment and sense of community on online social health support group},
  author={Sharma, Shwadhin and Khadka, Anita},
  journal={Information Technology \& People},
  volume={32},
  number={6},
  pages={1564--1590},
  year={2019},
  publisher={Emerald Publishing Limited}
}

@article{10.1145/3686898,
author = {Wang, Dennis and Eng, Jocelyn and Turpitka, Mykyta and Epstein, Daniel A.},
title = {Exploring Activity-Sharing Response Differences Between Broad-Purpose and Dedicated Online Social Platforms},
year = {2024},
journal = {Proc. CSCW},
month = nov
}

@inproceedings{de2018examination,
  title={An Examination of the Antecedents of Trust in Facebook Online Health Communities.},
  author={de Souza Tacco, Fabiana Martins and Sanchez, Ot{\'a}vio P and Connolly, Regina and Compeau, Deborah R},
  booktitle={ECIS},
  pages={150},
  year={2018}
}

@article{wang2018can,
  title={Can online social support be detrimental in stigmatized chronic diseases? A quadratic model of the effects of informational and emotional support on self-care behavior of HIV patients},
  author={Wang, Xunyi and Parameswaran, Srikanth and Bagul, Darshan Mahendra and Kishore, Rajiv},
  journal={JAMIA},
  year={2018},
  publisher={Oxford University Press}
}

@article{meng2021cancer,
  title={How cancer patients benefit from support networks offline and online: extending the model of structural-to-functional support},
  author={Meng, Jingbo and Rains, Steve A and An, Zheng},
  journal={Health Communication},
  year={2021},
  publisher={Taylor \& Francis}
}

@inproceedings{10.1145/3597638.3608400,
author = {Eagle, Tessa and Ringland, Kathryn E.},
title = {“You Can't Possibly Have ADHD”: Exploring Validation and Tensions around Diagnosis within Unbounded ADHD SMCs},
year = {2023},
booktitle = {Proc. ASSETS},
articleno = {29},
numpages = {17},
}

@inproceedings{10.1145/2531602.2531622,
author = {Massimi, Michael and Bender, Jackie L. and Witteman, Holly O. and Ahmed, Osman H.},
title = {Life transitions and online health communities: reflecting on adoption, use, and disengagement},
year = {2014},
booktitle = {Proc. CSCW},
location = {Baltimore, Maryland, USA},
}

@article{d2017social,
  title={Social networking online to recover from opioid use disorder: A study of community interactions},
  author={D’Agostino, Alexandra R and Optican, Allison R and Sowles, Shaina J and Krauss, Melissa J and Lee, Kiriam Escobar and Cavazos-Rehg, Patricia A},
  journal={Drug and alcohol dependence},
  volume={181},
  pages={5--10},
  year={2017},
  publisher={Elsevier}
}

@inproceedings{progga2023understanding,
  title={Understanding the online social support dynamics for postpartum depression},
  author={Progga, Farhat Tasnim and et al.},
  booktitle={Proc. CHI},
  pages={1--17},
  year={2023}
}

@article{10.1145/3402855,
author = {Smith, C. Estelle and Levonian, Zachary and Ma, Haiwei and et al.},
title = {"I Cannot Do All of This Alone": Exploring Instrumental and Prayer Support in Online Health Communities},
year = {2020},
issue_date = {October 2020},
publisher = {Association for Computing Machinery},
address = {New York, NY, USA},
issn = {1073-0516},
url = {https://doi.org/10.1145/3402855},
doi = {10.1145/3402855},
journal = {TOCHI},
month = aug,
articleno = {38},
numpages = {41}
}

@article{shah2022modeling,
  title={Modeling motivational interviewing strategies on an online peer-to-peer counseling platform},
  author={Shah, Raj Sanjay and Holt, Faye and Hayati, Shirley Anugrah and et al.},
  journal={Proc. CSCW},
  year={2022},
  publisher={ACM New York, NY, USA}
}

@article{10.1145/3449142,
author = {Naserianhanzaei, Elahe and Koschate-Reis, Miriam},
title = {Do Group Memberships Online Protect Addicts in Recovery Against Relapse? Testing the Social Identity Model of Recovery in the Online World},
year = {2021},
issue_date = {April 2021},
publisher = {Association for Computing Machinery},
address = {New York, NY, USA},
volume = {5},
number = {CSCW1},
url = {https://doi.org/10.1145/3449142},
doi = {10.1145/3449142},
journal = {Proc. ACM Hum.-Comput. Interact.},
month = apr,
articleno = {68},
numpages = {18}
}

@article{liu2017support,
  title={When support is needed: Social support solicitation and provision in an online alcohol use disorder forum},
  author={Liu, Yan and Kornfield, Rachel and Shaw, Bret R and et al.},
  journal={Digital health},
  volume={3},
  pages={2055207617704274},
  year={2017},
  publisher={SAGE Publications Sage UK: London, England}
}

@article{yao2022learning,
  title={Learning to become a volunteer counselor: Lessons from a peer-to-peer mental health community},
  author={Yao, Zheng and et al.},
  journal={CSCW},
  year={2022},
  publisher={ACM New York, NY, USA}
}

@article{berry2017whywetweetmh,
  title={\#WhyWeTweetMH: understanding why people use Twitter to discuss mental health problems},
  author={Berry, Natalie and Lobban, Fiona and Belousov, Maksim and Emsley, Richard and Nenadic, Goran and Bucci, Sandra},
  journal={JMIR},
  year={2017},
  publisher={JMIR Publications Toronto, Canada}
}

@article{10.1145/3479564,
author = {Papoutsaki, Alexandra and So, Samuel and Kenderova, Georgia and Shapiro, Bryan and Epstein, Daniel A.},
title = {Understanding Delivery of Collectively Built Protocols in an Online Health Community for Discontinuation of Psychiatric Drugs},
year = {2021},
issue_date = {October 2021},
publisher = {Association for Computing Machinery},
address = {New York, NY, USA},
volume = {5},
number = {CSCW2},
url = {https://doi.org/10.1145/3479564},
doi = {10.1145/3479564},
journal = {Proc. ACM Hum.-Comput. Interact.},
month = oct,
articleno = {420},
numpages = {29}
}

@inproceedings{10.1145/3411764.3445446,
author = {Peng, Zhenhui and Ma, Xiaojuan and Yang, Diyi and Tsang, Ka Wing and Guo, Qingyu},
title = {Effects of Support-Seekers’ Community Knowledge on Their Expressed Satisfaction with the Received Comments in Mental Health Communities},
year = {2021},
booktitle = {Proc. CHI},
articleno = {536},
numpages = {12},
location = {Yokohama, Japan},
}

@inproceedings{10.1145/2556288.2557108,
author = {Vlahovic, Tatiana A. and Wang, Yi-Chia and Kraut, Robert E. and Levine, John M.},
title = {Support matching and satisfaction in an online breast cancer support community},
year = {2014},
booktitle = {Proc. CHI},
location = {Toronto, Ontario, Canada},
}

@article{yan2018good,
  title={Good intentions, bad outcomes: The effects of mismatches between social support and health outcomes in an online weight loss community},
  author={Yan, Lu},
  journal={Production and Operations Management},
  year={2018},
  publisher={SAGE Publications Sage CA: Los Angeles, CA}
}

@article{abedin2019unpacking,
  title={Unpacking support types in online health communities: an application of attraction-selection-attrition theory},
  author={Abedin, Babak and Erfani, Shadi and Milne, David and Beattie, Annette and Fenerty, Kate},
  year={2019}
}

@inproceedings{10.1145/3290605.3300574,
author = {Yang, Diyi and Kraut, Robert E. and Smith, Tenbroeck and Mayfield, Elijah and Jurafsky, Dan},
title = {Seekers, Providers, Welcomers, and Storytellers: Modeling Social Roles in Online Health Communities},
year = {2019},
booktitle = {Proc. CHI},
location = {Glasgow, Scotland Uk},
}

@article{BUNTING2021108672,
title = {Socially-supportive norms and mutual aid of people who use opioids: An analysis of Reddit during the initial COVID-19 pandemic},
journal = {Drug and Alcohol Dependence},
year = {2021},
issn = {0376-8716},
doi = {https://doi.org/10.1016/j.drugalcdep.2021.108672},
url = {https://www.sciencedirect.com/science/article/pii/S0376871621001678},
author = {Amanda M. Bunting and David Frank and Joshua Arshonsky and Marie A. Bragg and Samuel R. Friedman and Noa Krawczyk}
}

@article{fan2018online,
  title={Online health communities: how do community members build the trust required to adopt information and form close relationships?},
  author={Fan, Hanmei and Lederman, Reeva},
  journal={EJIS},
  volume={27},
  number={1},
  pages={62--89},
  year={2018},
  publisher={Taylor \& Francis}
}

@article{10.1145/3434184,
author = {Levonian, Zachary and Dow, Marco and Erikson, Drew and et al.},
title = {Patterns of Patient and Caregiver Mutual Support Connections in an Online Health Community},
year = {2021},
issue_date = {December 2020},
journal = {Proc. CSCW},
month = jan,
articleno = {275},
numpages = {46}
}

@article{manga2022examining,
  title={Examining Users’ Information Disclosure and Audience Support Response Dynamics in OHCs: An Empirical Study},
  author={Manga, Joseph A and Andoh-Baidoo, Francis and Ayaburi, Emmanuel Wusuhon Yanibo and Escobari, Diego},
  year={2022}
}

@article{10.1145/3512938,
author = {Kim, Chelsea and Wang, Hao-Chuan},
title = {From Receivers to Givers: Understanding Practice of Reciprocity in an Online Support Community},
year = {2022},
issue_date = {April 2022},
publisher = {Association for Computing Machinery},
address = {New York, NY, USA},
volume = {6},
number = {CSCW1},
url = {https://doi.org/10.1145/3512938},
doi = {10.1145/3512938},
journal = {Proc. ACM Hum.-Comput. Interact.},
month = apr,
articleno = {91},
numpages = {17}
}

@article{rui2024provider,
  title={How do provider comm. strategies predict online patient satisfaction? A content analysis of online patient-provider comm. transcripts},
  author={Rui, Jian Raymond and Guo, Jieqiong and Yang, Keqing},
  journal={Digital Health},
  year={2024},
  publisher={SAGE Publications Sage UK: London, England}
}

@article{yao2022users,
  title={What users seek and share in online diabetes communities: examining similarities and differences in expressions and themes},
  author={Yao, Zhizhen and Zhang, Bin and Ni, Zhenni and Ma, Feicheng},
  journal={Aslib Journal of Information Management},
  volume={74},
  number={2},
  pages={311--331},
  year={2022},
  publisher={Emerald Publishing Limited}
}

@article{ortiz2023anyone,
  title={“Anyone else? Is this normal?”: anonymously seeking information on the Ovia Pregnancy App},
  author={Ortiz Juarez-Paz, Anna V and Doherty, Eileen F and Storch, Sharon L and Kallis, Rhiannon B and Kleinman, Steven B},
  journal={Atlantic Journal of Communication},
  volume={31},
  number={1},
  pages={30--51},
  year={2023},
  publisher={Taylor \& Francis}
}

@article{10.1145/3555133,
author = {Johnson, Jazette and Arnold, Vitica and Piper, Anne Marie and Hayes, Gillian R.},
title = {"It's a lonely disease": Cultivating Online Spaces for Social Support among People Living with Dementia and Dementia Caregivers},
year = {2022},
issue_date = {November 2022},
publisher = {Association for Computing Machinery},
address = {New York, NY, USA},
volume = {6},
number = {CSCW2},
url = {https://doi.org/10.1145/3555133},
doi = {10.1145/3555133},
journal = {Proc. ACM Hum.-Comput. Interact.},
month = nov,
articleno = {408},
numpages = {27}
}

@article{jiang2022effect,
  title={Effect of writing style on social support in online health communities: A theoretical linguistic analysis framework},
  author={Jiang, Shan and Liu, Xuan and Chi, Xiaotong},
  journal={Information \& Management},
  volume={59},
  number={6},
  pages={103683},
  year={2022},
  publisher={Elsevier}
}

@article{winker2000guidelines,
  title={Guidelines for medical and health information sites on the internet: principles governing AMA web sites},
  author={Winker, Margaret A and Flanagin, Annette and Chi-Lum, Bonnie and White, John and Andrews, Karen and Kennett, Robert L and DeAngelis, Catherine D and Musacchio, Robert A},
  journal={Jama},
  volume={283},
  number={12},
  pages={1600--1606},
  year={2000},
  publisher={American Medical Association}
}

@article{jong2021exchange,
  title={The exchange of informational support in online health communities at the onset of the COVID-19 pandemic: content analysis},
  author={Jong, Wesley and Liang, Ou Stella and Yang, Christopher C and others},
  journal={Jmirx med},
  year={2021},
  publisher={JMIR Publications Inc., Toronto, Canada}
}

@inproceedings{10.1145/3544548.3581489,
author = {Milton, Ashlee and Ajmani, Leah and DeVito, Michael Ann and Chancellor, Stevie},
title = {“I See Me Here”: Mental Health Content, Community, and Algorithmic Curation on TikTok},
year = {2023},
booktitle = {Proc. CHI},
articleno = {480},
numpages = {17},
location = {Hamburg, Germany},
}

@article{wang2021cass,
  title={Cass: Towards building a social-support chatbot for online health community},
  author={Wang, Liuping and Wang, Dakuo and Tian, Feng and Peng, Zhenhui and Fan, Xiangmin and Zhang, Zhan and Yu, Mo and Ma, Xiaojuan and Wang, Hongan},
  journal={Proc. CSCW},
  year={2021},
  publisher={ACM New York, NY, USA}
}

@article{ryan2009trust,
  title={Trust and participation in online usenet self-help communities},
  author={Ryan, Sherida},
  journal={International Journal of Self-Help \& Self-Care},
  volume={5},
  number={1},
  year={2009},
  publisher={SAGE Publications}
}

@article{james2022mediating,
  title={The mediating role of group dynamics in shaping received social support from active and passive use in online health communities},
  author={James, Tabitha L and Calderon, Eduardo D Villacis and B{\'e}langer, France and Lowry, Paul Benjamin},
  journal={Information \& Management},
  volume={59},
  number={3},
  pages={103606},
  year={2022},
  publisher={Elsevier}
}

@article{zhou2020exploring,
  title={Exploring the factors influencing consumers to voluntarily reward free health service contributors in online health communities: Empirical study},
  author={Zhou, Junjie and Liu, Fang and Zhou, Tingting},
  journal={JMIR},
  year={2020},
  publisher={JMIR Publications Toronto, Canada}
}

@article{10.1145/3614178.3614181,
author = {Huang, Kuang-Yuan and Long, Yoanna and Cui, Xiao},
title = {Let's Quit Together: Exploring Textual Factors Promoting Supportive Interactions in Online Cannabis Support Forums},
year = {2023},
issue_date = {August 2023},
publisher = {Association for Computing Machinery},
address = {New York, NY, USA},
volume = {54},
number = {3},
issn = {0095-0033},
url = {https://doi.org/10.1145/3614178.3614181},
doi = {10.1145/3614178.3614181},
journal = {SIGMIS Database},
month = aug,
pages = {11–36},
numpages = {26}
}

@article{10.1145/3415198,
author = {Johnson, Jazette and Black, Rebecca W. and Hayes, Gillian R.},
title = {Roles in the Discussion: An Analysis of Social Support in an Online Forum for People with Dementia},
year = {2020},
issue_date = {October 2020},
publisher = {Association for Computing Machinery},
address = {New York, NY, USA},
url = {https://doi.org/10.1145/3415198},
doi = {10.1145/3415198},
journal = {Proc. CSCW},
month = oct,
articleno = {127},
numpages = {30}
}

@article{10.1145/3631341.3631347,
author = {Fadel, Kelly J. and Jensen, Matthew L. and Meservy, Thomas O.},
title = {Online Information Filtering: The Role of Contextual Cues in Electronic Networks of Practice},
year = {2023},
issue_date = {November 2023},
publisher = {Association for Computing Machinery},
address = {New York, NY, USA},
volume = {54},
number = {4},
issn = {0095-0033},
url = {https://doi.org/10.1145/3631341.3631347},
doi = {10.1145/3631341.3631347},
journal = {SIGMIS Database},
month = oct,
pages = {77–106},
numpages = {30}
}

@article{liu2022reception,
  title={The reception of support in peer-to-peer online networks: Network position, support solicitation, and support provision in an online asthma caregivers group},
  author={Liu, Yan and Huang, Wensen and Luo, Dan},
  journal={Health informatics journal},
  volume={28},
  number={1},
  pages={14604582211066020},
  year={2022},
  publisher={SAGE Publications Sage UK: London, England}
}

@article{Choudhury2014MentalHD,
  title={Mental Health Discourse on Reddit: Self-Disclosure, Social Support, and Anonymity},
  author={Munmun {De Choudhury} and Sushovan De},
  journal={ICWSM},
  year={2014},
  url={https://api.semanticscholar.org/CorpusID:1578178}
}

@article{introne2020designing,
  title={Designing sustainable online support: Examining the effects of design change in 49 online health support communities},
  author={Introne, Joshua and Erickson, Ingrid and Semaan, Bryan and Goggins, Sean},
  journal={JAIST},
  year={2020},
  publisher={Wiley Online Library}
}

@article{imlawi2020understanding,
  title={Understanding the satisfaction and continuance intention of knowledge contribution by health professionals in online health communities},
  author={Imlawi, Jehad and Gregg, Dawn},
  journal={Informatics for Health and Social Care},
  volume={45},
  number={2},
  pages={151--167},
  year={2020},
  publisher={Taylor \& Francis}
}

@article{occhiogrosso2022feasibility,
  title={Feasibility of an online patient community to support older women with newly diagnosed breast cancer},
  author={Occhiogrosso, Rachel H and Ren, Siyang and Tayob, Nabihah and et al.},
  journal={Clinical Breast Cancer},
  year={2022},
  publisher={Elsevier}
}

@article{graham2017failure,
  title={The failure to increase social support: it just might be time to stop intervening (and start rigorously observing)},
  author={Graham, Amanda L and Papandonatos, George D and Zhao, Kang},
  journal={Translational behavioral medicine},
  volume={7},
  number={4},
  pages={816--820},
  year={2017},
  publisher={Oxford University Press}
}

@article{kepner2022types,
  title={Types and sources of stigma on opioid use treatment and recovery communities on reddit},
  author={Kepner, Wayne and Meacham, Meredith C and Nobles, Alicia L},
  journal={Substance use \& misuse},
  volume={57},
  number={10},
  pages={1511--1522},
  year={2022},
  publisher={Taylor \& Francis}
}

@article{10.1145/3449169,
author = {Allison, Kimberley R. and Patterson, Pandora and Guilbert, Daniel and Noke, Melissa and Husson, Olga},
title = {Logging On, Reaching Out, and Getting By: A Review of Self-reported Psychosocial Impacts of Online Peer Support for People Impacted by Cancer},
year = {2021},
issue_date = {April 2021},
publisher = {Association for Computing Machinery},
address = {New York, NY, USA},
url = {https://doi.org/10.1145/3449169},
doi = {10.1145/3449169},
journal = {Proc. CSCW},
month = apr,
articleno = {95},
numpages = {35}
}

@article{meyerhoff2022meeting,
  title={Meeting young adults' social support needs across the health behavior change journey: implications for digital mental health tools},
  author={Meyerhoff, Jonah and Kornfield, Rachel and et al.},
  journal={Proc. CSCW},
  year={2022},
  publisher={ACM New York, NY, USA}
}

@article{elsherief2021characterizing,
  title={Characterizing and identifying the prevalence of web-based misinformation relating to medication for opioid use disorder: machine learning approach},
  author={ElSherief, Mai and Sumner, Steven A and et al.},
  journal={JMIR},
  volume={23},
  number={12},
  pages={e30753},
  year={2021},
  publisher={JMIR Publications Toronto, Canada}
}

@inproceedings{park-etal-2024-valuescope,
    title = "{V}alue{S}cope: Unveiling Implicit Norms and Values via Return Potential Model of Social Interactions",
    author = "Park, Chan Young  and
      Li, Shuyue Stella  and
      Jung, Hayoung  and et al.",
    booktitle = "Findings of EMNLP",
    month = nov,
    year = "2024",
    url = "https://aclanthology.org/2024.findings-emnlp.972/",
    doi = "10.18653/v1/2024.findings-emnlp.972"
}

@inproceedings{laud2025large,
  title={Large-scale analysis of online questions related to opioid use disorder on reddit},
  author={Laud, Tanmay and Kacha-Ochana, Akadia and Sumner, Steven A and et al.},
  booktitle={Proc. ICWSM},
  year={2025}
}

@inproceedings{10.5555/3666122.3668142,
author = {Zheng, Lianmin and Chiang, Wei-Lin and Sheng, Ying and Zhuang, Siyuan and Wu, Zhanghao and Zhuang, Yonghao and Lin, Zi and Li, Zhuohan and Li, Dacheng and Xing, Eric P. and Zhang, Hao and Gonzalez, Joseph E. and Stoica, Ion},
title = {Judging LLM-as-a-judge with MT-bench and Chatbot Arena},
year = {2023},
booktitle = {Proc. NeurIPS},
articleno = {2020},
numpages = {29},
location = {New Orleans, LA, USA},
}

@article{dubois2023alpacafarm,
  title={Alpacafarm: A simulation framework for methods that learn from human feedback},
  author={Dubois, Yann and Li, Chen Xuechen and et al.},
  journal={NeurIPS},
  year={2023}
}

@article{ziems-etal-2024-large,
    title = "Can Large Language Models Transform Computational Social Science?",
    author = "Ziems, Caleb  and
      Held, William  and
      Shaikh, Omar  and
      Chen, Jiaao  and
      Zhang, Zhehao  and
      Yang, Diyi",
    journal = "Computational Linguistics",
    volume = "50",
    number = "1",
    month = mar,
    year = "2024",
    address = "Cambridge, MA",
    publisher = "MIT Press",
    url = "https://aclanthology.org/2024.cl-1.8/",
    doi = "10.1162/coli_a_00502",
    pages = "237--291"
}

@misc{cheng2025elephantmeasuringunderstandingsocial,
      title={ELEPHANT: Measuring and understanding social sycophancy in LLMs}, 
      author={Myra Cheng and Sunny Yu and Cinoo Lee and Pranav Khadpe and Lujain Ibrahim and Dan Jurafsky},
      year={2025},
      eprint={2505.13995},
      archivePrefix={arXiv},
      primaryClass={cs.CL},
      url={https://arxiv.org/abs/2505.13995}, 
}

@article{Mittal_Jung_ElSherief_Mitra_DeChoudhury_2025, title={Online Myths on Opioid Use Disorder: A Comparison of Reddit and Large Language Model}, url={https://ojs.aaai.org/index.php/ICWSM/article/view/35870}, DOI={10.1609/icwsm.v19i1.35870}, abstractNote={Online communities on Reddit are a popular choice among people with opioid use disorder (OUD) to seek information on drug use, withdrawal symptoms, and recovery. LLM-powered chatbots (e.g., ChatGPT) are widely being adopted as question-answer systems for health-related queries. However, such online health information seeking could potentially be hindered by myths and misinformation on OUD, misleading or causing genuine harm to people with OUD. In this work, we examine the prevalence of 5 OUD-related myths, on treatment models and patient characteristics, within human- (taken from Reddit) and LLM-generated responses to queries on OUD. We further explore the framing strategies used within responses (both human- and LLM-generated) promoting and countering the myths. We found that all 5 myths were more widespread within human-generated responses. In addition, myth-promoting responses adopted trustworthy and authoritative framings, compared to knowledge-imparting linguistic cues within those countering the myths. Our work offers recommendations to reduce online OUD misinformation.}, journal={Proc. ICWSM}, author={Mittal, Shravika and Jung, Hayoung and ElSherief, Mai and Mitra, Tanushree and De Choudhury, Munmun}, year={2025}, month={Jun.}, }

@article{10.1145/3757445,
author = {Lloyd, Travis and Reagle, Joseph and Naaman, Mor},
title = { `There Has To Be a Lot That We're Missing': Moderating AI-Generated Content on Reddit},
year = {2025},
journal = {Proc. CSCW},
month = oct,
numpages = {24}
}

@inproceedings{10.1145/3706598.3713292,
author = {Lloyd, Travis and Gosciak, Jennah and Nguyen, Tung and Naaman, Mor},
title = {AI Rules? Characterizing Reddit Community Policies Towards AI-Generated Content},
year = {2025},
booktitle = {Proc. CHI},
articleno = {9},
numpages = {19},
location = {
},
}

@article{Balsamo_Bajardi_DeFrancisciMorales_Monti_Schifanella_2023, title={The Pursuit of Peer Support for Opioid Use Recovery on Reddit}, volume={17}, url={https://ojs.aaai.org/index.php/ICWSM/article/view/22122}, DOI={10.1609/icwsm.v17i1.22122}, abstractNote={Individuals suffering from Opioid Use Disorder and other socially stigmatized conditions often rely on peer support groups to find comfort and motivation while treating their condition. Many may face barriers in accessing peer support treatment, such as shame and social stigma, seclusion, or mobility restrictions. In this study, we quantitatively characterize the potential of the Reddit community in offering these individuals an online alternative to receiving peer support. By analyzing the social interactions of thousands of users during the start of opioid use recovery, we uncover that a particular Reddit community exhibits many characteristics similar to in-person peer support groups, featuring the exchange of support, trust, status, and similar experiences. We find that the supportive behavior of this community nudges users to change their personal behavior, and promotes abandoning opioid-related communities in favor of recovery-oriented relationships. Finally, we find that recognition, acknowledgment, and knowledge exchange are the most relevant factors in sustained engagement with the recovery community. Given this evidence, we suggest that this online community may constitute a complement or a surrogate to peer support groups when in-person meetings are not desirable or possible. Our work might inspire harm reduction policies and interventions to favor successful rehabilitation and is fundamental for future research about the use of digital media for recovery support.}, number={1}, journal={Proc. ICWSM}, author={Balsamo, Duilio and Bajardi, Paolo and De Francisci Morales, Gianmarco and Monti, Corrado and Schifanella, Rossano}, year={2023}, month={Jun.}, pages={12-23} }

@misc{openai2024gpt4technicalreport,
      title={GPT-4 Technical Report}, 
      author={OpenAI and Josh Achiam and Steven Adler and {et al.}},
      year={2024},
      eprint={2303.08774},
      archivePrefix={arXiv},
      primaryClass={cs.CL},
      url={https://arxiv.org/abs/2303.08774}, 
}

@misc{openaiOpenAIPlatform,
	author = {OpenAI},
	title = {Prompt engineering},
	howpublished = {\url{https://platform.openai.com/docs/guides/prompt-engineering/prompt-engineering}},
	year = {2023},
}

@article{kiesler2012regulating,
  title={Regulating behavior in online communities},
  author={Kiesler, Sara and Kraut, Robert and Resnick, Paul and Kittur, Aniket},
  journal={Building successful online communities: Evidence-based social design},
  volume={1},
  pages={4--2},
  year={2012},
  publisher={MIT Press Cambridge, MA}
}

@article{burnett2001small,
  title={Small worlds: Normative behavior in virtual communities and feminist bookselling},
  author={Burnett, Gary and Besant, Michele and Chatman, Elfreda A},
  journal={JAIST},
  volume={52},
  number={7},
  pages={536},
  year={2001},
  publisher={Wiley Periodicals Inc.}
}

@inproceedings{arguello2006talk,
  title={Talk to me: foundations for successful individual-group interactions in online communities},
  author={Arguello, Jaime and Butler, Brian S and Joyce, Elisabeth and Kraut, Robert and Ling, Kimberly S and Ros{\'e}, Carolyn and Wang, Xiaoqing},
  booktitle={Proc. CHI},
  year={2006}
}

@article{Fiesler_Jiang_McCann_Frye_Brubaker_2018, title={Reddit Rules! Characterizing an Ecosystem of Governance}, volume={12}, url={https://ojs.aaai.org/index.php/ICWSM/article/view/15033}, DOI={10.1609/icwsm.v12i1.15033}, abstractNote={ &lt;p&gt; The social sharing and news aggregation site Reddit provides a unique example of an ecosystem of community-created rules. Not only do individual subreddits create and enforce their own regulations, but site-wide guidelines and norms may also influence behavior. This paper reports on a mixed-methods study of 100,000 subreddits and their rules. Our findings characterize the types of rules across Reddit, the frequency of rules at scale, and patterns of rules based on subreddit characteristics. We find that rules appear to be context-dependent for individual subreddits but also share common characteristics across the site. Taken together, our findings provide a rich description of this ecosystem of rules, motivating further inquiry into underlying mechanisms for rule formation and enforcement in online communities. &lt;/p&gt; }, number={1}, journal={Proc. ICWSM}, author={Fiesler, Casey and Jiang, Jialun and McCann, Joshua and et al.}, year={2018}, month={Jun.} }

@article{10.1145/3274301,
author = {Chandrasekharan, Eshwar and Samory, Mattia and Jhaver, Shagun and et al.},
title = {The Internet's Hidden Rules: An Empirical Study of Reddit Norm Violations at Micro, Meso, and Macro Scales},
year = {2018},
issue_date = {November 2018},
publisher = {Association for Computing Machinery},
address = {New York, NY, USA},
url = {https://doi.org/10.1145/3274301},
doi = {10.1145/3274301},
journal = {Proc. CSCW},
month = nov,
articleno = {32},
numpages = {25}
}

@article{thomas2003general,
  title={A general inductive approach for qualitative data analysis},
  author={Thomas, David R},
  year={2003},
  publisher={Auckland, New Zealand}
}

@misc{nihWordsMatter,
	author = {{NIDA}},
	title = {{W}ords {M}atter - {T}erms to {U}se and {A}void {W}hen {T}alking {A}bout {A}ddiction},
	howpublished = {\url{https://nida.nih.gov/nidamed-medical-health-professionals/health-professions-education/words-matter-terms-to-use-avoid-when-talking-about-addiction}},
	year = {2021},
}

@misc{nihWordsMatter_2,
	author = {{NIDA}},
	title = {{P}referred {L}anguage for {T}alking {A}bout {A}ddiction},
	howpublished = {\url{https://nida.nih.gov/research-topics/addiction-science/words-matter-preferred-language-talking-about-addiction}},
	year = {2021},
}

@inproceedings{10.1145/3544548.3581318,
author = {Zhou, Jiawei and Zhang, Yixuan and Luo, Qianni and et al.},
title = {Synthetic Lies: Understanding AI-Generated Misinformation and Evaluating Algorithmic and Human Solutions},
year = {2023},
booktitle = {Proc. CHI},
articleno = {436},
numpages = {20},
location = {Hamburg, Germany},
}

@article{10.1145/3687030,
author = {Agarwal, Dhruv and Shahid, Farhana and Vashistha, Aditya},
title = {Conversational Agents to Facilitate Deliberation on Harmful Content in WhatsApp Groups},
year = {2024},
issue_date = {November 2024},
publisher = {Association for Computing Machinery},
address = {New York, NY, USA},
volume = {8},
number = {CSCW2},
url = {https://doi.org/10.1145/3687030},
doi = {10.1145/3687030},
journal = {Proc. ACM Hum.-Comput. Interact.},
month = nov,
articleno = {491},
numpages = {32}
}

@misc{mittal2025exposurecontentwrittenlarge,
      title={Exposure to Content Written by Large Language Models Can Reduce Stigma Around Opioid Use Disorder in Online Communities}, 
      author={Shravika Mittal and Darshi Shah and Shin Won Do and Mai ElSherief and Tanushree Mitra and Munmun De Choudhury},
      year={2025},
      eprint={2504.10501},
      archivePrefix={arXiv},
      primaryClass={cs.SI},
      url={https://arxiv.org/abs/2504.10501}, 
}

@misc{techcrunchInstagramIntroduces,
	author = {Ivan Mehta},
	title = {{I}nstagram introduces {G}en{A}{I} powered background editing tool},
	howpublished = {\url{https://techcrunch.com/2023/12/14/instagram-introduces-gen-ai-powered-background-editing-tool/}},
	year = {2023},
}

@inproceedings{10.1145/2858036.2858356,
author = {Kiene, Charles and Monroy-Hern\'{a}ndez, Andr\'{e}s and Hill, Benjamin Mako},
title = {Surviving an "Eternal September": How an Online Community Managed a Surge of Newcomers},
year = {2016},
booktitle = {Proc. CHI},
numpages = {5},
location = {San Jose, California, USA},
}

@inproceedings{10.1145/2998181.2998277,
author = {Seering, Joseph and Kraut, Robert and Dabbish, Laura},
title = {Shaping Pro and Anti-Social Behavior on Twitch Through Moderation and Example-Setting},
year = {2017},
booktitle = {Proc. CSCW},
location = {Portland, Oregon, USA},
}

@article{lampe2014crowdsourcing,
  title={Crowdsourcing civility: A natural experiment examining the effects of distributed moderation in online forums},
  author={Lampe, Cliff and Zube, Paul and Lee, Jusil and et al.},
  journal={Government Information Quarterly},
  year={2014},
  publisher={Elsevier}
}

@inproceedings{park-etal-2021-detecting-community,
    title = "Detecting Community Sensitive Norm Violations in Online Conversations",
    author = "Park, Chan Young  and
      Mendelsohn, Julia  and
      Radhakrishnan, Karthik  and
      et al.",
    booktitle = "Findings of EMNLP",
    month = nov,
    year = "2021"
}

@article{10.1145/3757474,
author = {Guo, Qingyu and Zhang, Yuqi and Yuan, Kangyu and et al.},
title = {Exploring the Evolvement of User Engagement in Online Creative Community under the Surge of Generative AI},
year = {2025},
issue_date = {November 2025},
publisher = {Association for Computing Machinery},
address = {New York, NY, USA},
url = {https://doi.org/10.1145/3757474},
doi = {10.1145/3757474},
journal = {Proc. CSCW},
month = oct,
articleno = {CSCW293},
numpages = {35}
}

@inproceedings{thakur-etal-2025-judging,
    title = "Judging the Judges: Evaluating Alignment and Vulnerabilities in {LLM}s-as-Judges",
    author = "Thakur, Aman Singh  and
      Choudhary, Kartik  and
      Ramayapally, Venkat Srinik  and
      Vaidyanathan, Sankaran  and
      Hupkes, Dieuwke",
    booktitle = "Proceedings of the Fourth Workshop on Generation, Evaluation and Metrics (GEM{\texttwosuperior})",
    month = jul,
    year = "2025",
    url = "https://aclanthology.org/2025.gem-1.33/"
}

@incollection{nagler2017conflicting,
  title={Conflicting information and message competition in health and risk messaging},
  author={Nagler, Rebekah H and LoRusso, Susan M},
  booktitle={Oxford research encyclopedia of communication},
  year={2017}
}

@inproceedings{kelly2025understanding,
  title={Understanding Gender Bias in AI-Generated Product Descriptions},
  author={Kelly, Markelle and Tahaei, Mohammad and Smyth, Padhraic and Wilcox, Lauren},
  booktitle={Proc. FAccT},
  pages={2587--2615},
  year={2025}
}

@article{10.1145/3757544,
author = {Zhou, Jiawei and Chen, Amy Z. and Shah, Darshi and Schwab-Reese, Laura M. and De Choudhury, Munmun},
title = {A Risk Taxonomy and Reflection Tool for Large Language Model Adoption in Public Health},
year = {2025},
issue_date = {November 2025},
journal = {Proc. CSCW},
month = oct
}

@online{handa2025education,
author = {Kunal Handa and Drew Bent and Alex Tamkin and Miles McCain and Esin Durmus and Michael Stern and Mike Schiraldi and Saffron Huang and Stuart Ritchie and Steven Syverud and Kamya Jagadish and Margaret Vo and Matt Bell and Deep Ganguli},
title = {Anthropic Education Report: How University Students Use Claude},
date = {2025-04-08},
year = {2025},
url = {https://www.anthropic.com/news/anthropic-education-report-how-university-students-use-claude},
}

@inproceedings{10.1145/3582269.3615599,
author = {Kotek, Hadas and Dockum, Rikker and Sun, David},
title = {Gender bias and stereotypes in Large Language Models},
year = {2023},
booktitle = {Proc. CI},
location = {Delft, Netherlands},
}

@article{gebru2021datasheets,
  title={Datasheets for datasets},
  author={Gebru, Timnit and Morgenstern, Jamie and Vecchione, Briana and Vaughan, Jennifer Wortman and Wallach, Hanna and Iii, Hal Daum{\'e} and Crawford, Kate},
  journal={Communications of the ACM},
  volume={64},
  number={12},
  pages={86--92},
  year={2021},
  publisher={ACM New York, NY, USA}
}

@misc{fair,
    title="The FAIR Data principles",
year = 2020,
    author="{FORCE11}",
howpublished="\url{https://force11.org/info/the-fair-data-principles/}"
}

\section{Paper Checklist}

\begin{enumerate}

\item For most authors...
\begin{enumerate}
    \item  Would answering this research question advance science without violating social contracts, such as violating privacy norms, perpetuating unfair profiling, exacerbating the socio-economic divide, or implying disrespect to societies or cultures?
    \answerYes{Yes, as described in the Introduction, this work advances OHCs and support environments by clarifying how AI-generated content interacts with community norms, the risks it introduces, and how moderation practices can be adapted to regulate its use.}
  \item Do your main claims in the abstract and introduction accurately reflect the paper's contributions and scope?
    \answerYes{Yes.}
   \item Do you clarify how the proposed methodological approach is appropriate for the claims made? 
    \answerYes{Yes, refer to Methods.}
   \item Do you clarify what are possible artifacts in the data used, given population-specific distributions?
    \answerYes{Yes, refer to Data.}
  \item Did you describe the limitations of your work?
    \answerYes{Yes, refer to Limitations and Future Work.}
  \item Did you discuss any potential negative societal impacts of your work?
    \answerYes{Yes, refer to the Ethics Statement below.}
      \item Did you discuss any potential misuse of your work?
    \answerNA{NA}
    \item Did you describe steps taken to prevent or mitigate potential negative outcomes of the research, such as data and model documentation, data anonymization, responsible release, access control, and the reproducibility of findings?
    \answerYes{Yes, refer to the Ethics Statement below.}
  \item Have you read the ethics review guidelines and ensured that your paper conforms to them?
    \answerYes{Yes.}
\end{enumerate}

\item Additionally, if your study involves hypotheses testing...
\begin{enumerate}
  \item Did you clearly state the assumptions underlying all theoretical results?
    \answerNA{NA}
  \item Have you provided justifications for all theoretical results?
    \answerNA{NA}
  \item Did you discuss competing hypotheses or theories that might challenge or complement your theoretical results?
    \answerNA{NA}
  \item Have you considered alternative mechanisms or explanations that might account for the same outcomes observed in your study?
    \answerNA{NA}
  \item Did you address potential biases or limitations in your theoretical framework?
    \answerNA{NA}
  \item Have you related your theoretical results to the existing literature in social science?
    \answerNA{NA}
  \item Did you discuss the implications of your theoretical results for policy, practice, or further research in the social science domain?
    \answerNA{NA}
\end{enumerate}

\item Additionally, if you are including theoretical proofs...
\begin{enumerate}
  \item Did you state the full set of assumptions of all theoretical results?
    \answerNA{NA}
	\item Did you include complete proofs of all theoretical results?
    \answerNA{NA}
\end{enumerate}

\item Additionally, if you ran machine learning experiments...
\begin{enumerate}
  \item Did you include the code, data, and instructions needed to reproduce the main experimental results (either in the supplemental material or as a URL)?
    \answerYes{Yes, we (1) provide a source for the Reddit-QA  dataset~\cite{laud2025large}, (2) describe the procedures used to generate the AI responses (refer to Data), and (3) detail the annotation and evaluation process (refer to Methods).}
  \item Did you specify all the training details (e.g., data splits, hyperparameters, how they were chosen)?
    \answerYes{Yes, we specify training details for the LLM judges in Methods and provide sample judge prompts in the Appendix.}
     \item Did you report error bars (e.g., with respect to the random seed after running experiments multiple times)?
    \answerNA{NA}
	\item Did you include the total amount of compute and the type of resources used (e.g., type of GPUs, internal cluster, or cloud provider)?
    \answerNo{No, because our work was not compute intensive and we did not use any external storage or compute resources.}
     \item Do you justify how the proposed evaluation is sufficient and appropriate to the claims made? 
    \answerYes{Yes, refer to Methods.}
     \item Do you discuss what is ``the cost`` of misclassification and fault (in)tolerance?
    \answerYes{Yes. For our LLM judges, we report performance metrics (Table~\ref{tab:performance}) and provide an error analysis describing misclassifications (in the Appendix).}
  
\end{enumerate}

\item Additionally, if you are using existing assets (e.g., code, data, models) or curating/releasing new assets, \textbf{without compromising anonymity}...
\begin{enumerate}
  \item If your work uses existing assets, did you cite the creators?
    \answerYes{Yes, refer to Data.}
  \item Did you mention the license of the assets?
    \answerNA{NA}
  \item Did you include any new assets in the supplemental material or as a URL?
    \answerNA{NA}
  \item Did you discuss whether and how consent was obtained from people whose data you're using/curating?
    \answerNo{No, because we utilized a publicly available, retrospective dataset~\cite{laud2025large}. Working with this data did not constitute human subjects research and thus, informed consent was not required from the authors of the original posts. However, we followed best practices, worked with deidentified data and refrained from revealing any identifiable markers to avoid potential harm to those who authored or were referenced in the original posts (see Ethics Statement below).}
  \item Did you discuss whether the data you are using/curating contains personally identifiable information or offensive content?
    \answerYes{Yes, refer to the Ethics Statement below.}
\item If you are curating or releasing new datasets, did you discuss how you intend to make your datasets FAIR (see \citet{fair})?
\answerNA{NA}
\item If you are curating or releasing new datasets, did you create a Datasheet for the Dataset (see \citet{gebru2021datasheets})? 
\answerNA{NA}
\end{enumerate}

\item Additionally, if you used crowdsourcing or conducted research with human subjects, \textbf{without compromising anonymity}...
\begin{enumerate}
  \item Did you include the full text of instructions given to participants and screenshots?
    \answerNA{NA}
  \item Did you describe any potential participant risks, with mentions of Institutional Review Board (IRB) approvals?
    \answerNo{No. We used a publicly available dataset and did not interact with the authors of the original posts or the individuals referenced in them. As an observational study of retrospectively collected data, our research did not constitute human subjects research under our IRB guidelines, and thus formal IRB approval was not required. Nevertheless, ethical considerations in web-based research extend beyond IRB oversight. To mitigate potential risks, we used deidentified data and included only paraphrased excerpts in the paper to prevent traceability and minimize potential harm to the original authors (see Ethics Statement below).}
  \item Did you include the estimated hourly wage paid to participants and the total amount spent on participant compensation?
    \answerNA{NA}
   \item Did you discuss how data is stored, shared, and deidentified?
   \answerYes{Yes, refer to the Ethics Statement below.}
\end{enumerate}

\end{enumerate}

\section{Ethics Statement}
Following best practices~\cite{chancellor2019human}, we worked with deidentified publicly available data, siloed servers only accessible to the authors of this work, and refrained from sharing raw and personally identifiable data in any form (e.g., with annotators or during the use of language model APIs). All referenced excerpts taken from the Reddit-QA dataset were paraphrased to reduce traceability and potential harm to those who authored (or were referred to in) the original posts. We note that the annotation-intensive methodology required to conduct this work may have imposed substantial emotional labor on our human raters or exposed them to distressing content (given the sensitive nature of the study). Nonetheless, annotators were encouraged to take frequent breaks, pause or discontinue annotation as needed, and prioritize their well-being. The authors declare no competing interests.

\setcounter{table}{0}
\renewcommand{\thetable}{A\arabic{table}}
\setcounter{figure}{0}  
\renewcommand{\thefigure}{A\arabic{figure}}
\section{Appendix}

\subsection{Prompt}
We queried GPT-4 with a carefully designed prompt (see Table~\ref{tab:gpt4-prompt}) based on existing guidelines~\cite{openaiOpenAIPlatform}. 
Specifically, we instructed GPT-4 to respond to a support-seeking query. The prompt included the original post containing the query, taken as is from Reddit-QA, along with additional context, i.e., the name and description of the subreddit from where the post was sourced.

\begin{table}[!h]
    \centering
    \begin{tabular}{@{}p{\columnwidth}@{}}
    \toprule
    You are familiar with different online health communities, such as health-related subreddits on Reddit. In these communities, support is often provided in response to posts seeking help. Given a support-seeking post (POST), taken from \textsc{[subreddit name]}, generate an appropriate response, which could be offered to help the original poster on the subreddit. For context, the \textsc{[subreddit name]} is described as \textsc{subreddit description}.\\
    
    POST: \textsc{[post from Reddit-QA]} \\
    \bottomrule
    \end{tabular}
    \caption{Prompt used to instruct GPT-4}
    \label{tab:gpt4-prompt}
\end{table}

\begin{table*}[!th]
    \centering
    \small{
    \begin{tabular}{@{}p{0.22\textwidth}|p{0.7\textwidth}@{}}
    \toprule
        \textbf{Category} & \textbf{Keywords} \\
    \midrule
        Implicit norms & implicit norm(s), implicit rule(s), guideline(s), expectation(s), preference(s)  \\ \hline
        Support provision & support provision, provide support, offer support, provide social support, offer social support\\ \hline
        Online health community & online health community, online health platform, online health forum \\
    \bottomrule
    \end{tabular}}
            \caption{Keywords used to identify scholarly articles describing implicit norms governing support provision in OHCs. We used an `OR' operator within a category and an `AND' operator across to query the databases.}
    \label{tab:keywords}
\end{table*}

\subsection{Scoping Review Procedure}
We queried three search engines, ACM Digital Library, Google Scholar (using the Publish or Perish software), and Web of Science, to consider a broad range of conference proceedings and journal articles. 
The scoping study was guided by an existing conceptualization of implicit norms, i.e., unwritten rules or guidelines that define desirable, community-preferred attributes of a response providing support~\cite{burnett2003beyond}. We referred to this conceptualization, prior review studies on OHCs~\cite{carron2015help,10.1145/3461778.3462100}, and influential papers on online support provision~\cite{andalibi2018social} to curate a list of keywords (see Table~\ref{tab:keywords}) centered on: implicit norms governing support provision in OHCs.
Using a combination of these, with an `OR' operator within a category and an `AND' operator across, we searched for relevant literature in August '25 for articles published between January '08 and August '25, overlapping with the emergence of academic research on OHCs~\cite{10.1145/3461778.3462100}. 

\subsection{List of Relevant Articles}
Table~\ref{tab:references} lists the 61 relevant articles identified through a rapid literature review, along with the implicit norms governing support provision in OHCs discussed in each paper.

\begin{table*}[!ht]
    \centering
    \begin{tabular}{@{}p{0.24\textwidth}|p{0.75\textwidth}@{}}
    \toprule
       \textbf{Implicit Norm}  & \textbf{References} \\
    \midrule
        Cite sources & \citet{fan2018online,winker2000guidelines,progga2023understanding,10.1145/3597638.3608400,jong2021exchange,jiang2022effect,graham2017failure,10.1145/3173574.3174240,10.1145/3461778.3462100,10.1145/3415198,liu2022reception} \\ \hline

        Respect anonymity & \citet{andalibi2018social,progga2023understanding,Choudhury2014MentalHD,imlawi2020understanding,occhiogrosso2022feasibility,10.1145/3614178.3614181,10.1145/3449142} \\ \hline

        Be accountable & \citet{10.1145/3479564} \\ \hline
        
        Use coded language & \citet{10.1145/3173574.3174240,10.1145/3173574.3174215,Saha_Sharma_2020,althoff-etal-2016-large,10.1145/2998181.2998243,10.1145/3290605.3300354,cebbb012-2239-30e3-b067-1101aabbd5ab} \\ \hline

        Use simple language & \citet{wang2021cass,rui2024provider,ryan2009trust,james2022mediating,abedin2019unpacking,wang2018can,zhou2020exploring,10.1145/3614178.3614181,10.1145/3461778.3462100,10.1145/3415198} \\ \hline

        Avoid a limited response & \citet{10.1145/3631341.3631347} \\ \hline

        Use appropriate communication strategies & \citet{althoff-etal-2016-large} \\ \hline
        
        Use a dynamic writing style & \citet{Saha_Sharma_2020} \\ \hline

        Provide well-structured responses & \citet{10.1145/3544548.3581489} \\ \hline
        
        Avoid information overload & \citet{MPINGANJIRA2018686,zhang2013facebook,sharma2019role,10.1145/3686898,de2018examination,wang2018can,meng2021cancer,10.1145/3597638.3608400,10.1145/2531602.2531622} \\ \hline
        
        Validate, affirm, normalize & \citet{d2017social,progga2023understanding,10.1145/3597638.3608400,10.1145/3402855,shah2022modeling,10.1145/3449142,liu2017support,yao2022learning,10.1145/3290605.3300574,10.1145/2531602.2531622,zhang2013facebook,althoff-etal-2016-large} \\ \hline 
        
        Instill hope & \citet{d2017social,berry2017whywetweetmh} \\ \hline
        
        Reflect & \citet{shah2022modeling} \\ \hline

        Reduce isolation & \citet{10.1145/2998181.2998243,berry2017whywetweetmh,yao2022learning,10.1145/3597638.3608400} \\ \hline
        
        Combat stigma & \citet{berry2017whywetweetmh,10.1145/3479564,10.1145/3597638.3608400,progga2023understanding,10.1145/2998181.2998243,berry2017whywetweetmh,jong2021exchange,d2017social,ortiz2023anyone} \\ \hline
        
        Support matching & \citet{10.1145/3411764.3445446,10.1145/2556288.2557108,yan2018good,abedin2019unpacking} \\ \hline

        Personalize support & \citet{kepner2022types,10.1145/3544548.3581489,10.1145/3449169,liu2017support,meyerhoff2022meeting,10.1145/3402855,10.1145/3411764.3445446} \\ \hline

        Share lived experience & \citet{10.1145/3290605.3300574,BUNTING2021108672,fan2018online,shah2022modeling,10.1145/3434184,10.1145/3479564,yao2022learning,manga2022examining,10.1145/3512938,meng2021cancer,rui2024provider,yao2022users,ortiz2023anyone,10.1145/3555133,jiang2022effect,10.1145/3597638.3608400,progga2023understanding} \\ \hline

        Remember recent interactions & \citet{fan2018online} \\ \hline
        
        Build long-term trust & \citet{fan2018online} \\ \hline
        
        Facilitate long-term interaction & \citet{BUNTING2021108672,10.1145/2531602.2531622,introne2020designing} \\ 
    \bottomrule
    \end{tabular}
    \caption{References to relevant articles identified during the rapid literature review, which mention implicit norms governing support provision in OHCs.}
    \label{tab:references}
\end{table*}

\subsection{Implicit Norms Inferred Through Open-Coding}
Through an open-coding approach, human annotators surfaced that high-rated comments typically contained a \emph{positive tone}. They were \emph{validating}, \emph{non-judgmental to the support seeker}, and \emph{instilled hope}. Many elicited a \emph{shared lived experience} in order to affirm the seekers' concerns and provide support. In a few, \emph{support providers identified themselves as a trusted expert}, e.g., as a healthcare provider or someone with a lived experience to build credibility. Additionally, many high-rated comments sought to \emph{combat stigma}, e.g., by supporting the use of medications for opioid use disorder (MOUD). They often used \emph{non-technical language}, \emph{community-specific phrases} (e.g., `subs' to refer to suboxone), and an \emph{easy-to-consume formatting} (e.g., bulleted lists, step-by-step responses). They further \emph{responded directly}, \emph{matching the seekers' needs.}

In contrast, coders noted that low-rated comments were \emph{discouraging}. Comments that \emph{reinforced stigma} related to MOUD were typically low-rated, e.g., those that advised quitting opioids cold turkey or suggested alternative substances to MOUD to manage withdrawal (e.g., Kratom). Some low-rated comments shared lived experiences; however, they often focused on \emph{venting personal frustrations} rather than attempting to build a connection or provide support to the seeker, thereby reinforcing barriers to care and recovery. Several low-rated comments were \emph{very short} or \emph{provided limited information}, while others led to \emph{cognitive overload} including broad, round-about phrases with limited clarity, repeated information, or information tangential to seekers' needs. 
Table~\ref{tab:implicit-norms} presents example posts and corresponding high- and low-rated comments reviewed by annotators, along with the implicit norms inferred through open-coding.

\begin{table*}[t]
    \begin{center}
    \begin{tabular}{@{}p{0.3\textwidth}|p{0.3\textwidth}|p{0.3\textwidth}@{}}
    \toprule
    \textbf{Original Post} & \textbf{High-Rated Comment} & \textbf{Low-Rated Comment} \\
    \midrule
    Does anyone ever regret opting to go the methadone maintenance route, rather than just quitting cold-turkey? Because I do. [...] & I was homeless, miserable [...]. \hl{(shares lived experience)} Methadone has allowed me to not only get my life to a manageable place, but it has given me a chance to start over again. \hl{(combats stigma)} [...]
    I've got 4 cats and a dog, my own apartment. My life has gotten so much better. All of the suffering and hard work has been worth it. \hl{(instills hope)} & Yes, biggest mistake of my life. \hl{(reinforces stigma; provides limited information)} \\ \hline
    
    In my group meetings, every now and then someone admits they relapsed but is sober again. They still get their prescription and aren't discharged. Is it really that relaxed? Can you just admit you used and keep getting your treatment? & I own and operate a clinic. My medical director and I decide on all discharges together. \hl{(identifies as a trusted expert)} Relapse happens, and discharging a patient may set them up for failure. Self-reporting a relapse is a good sign. \hl{(uses simple, non-technical language)} [...] Patients need support more than ever at that point. \hl{(uses positive, non-judgmental tone)} & [...]  I have messaged a bunch of times about not being able to do those hours. They just don't get it. \hl{(provides an off-topic response; vents personal frustration)} Stressed, missing events because we have ``the flu'', waking up with it on our minds, going to sleep the same way, feeling pissed about what could have been [...] \hl{(overloads with unnecessary detail)} \\ \hline
    
    I have got to get this off my chest. Being on suboxone has been a blessing in my life [...]. I have had to deal with the pharmacy judging me. The very people that should understand, the ones that take suboxone have also shamed me for being on it. [...] & Everyone who tries shaming you for taking subs is a piece of shit. \hl{(uses community-specific terms)} If what you're doing is keeping you from that terrible way of life then whatever you're doing isn't wrong. \hl{(provides validation)} So don't listen to anyone's ``expert opinion''. If it wasn't for subs I'd either be dead or in prison. \hl{(shares lived experience)} [...] & You don't need to be on suboxone for years. That is a weakness. \hl{(uses negative, discouraging tone; reinforces stigma)} I was on them for 5 years. My doctor tried to keep me on them. It took me 5 months, but I quit! \hl{(vents personal frustration)} \\ \hline
    
    For people on Suboxone maintenance: does ``less is more'' really work, or does it just lead to cravings and boredom like some say? & From what I've seen, ``less is more'' really depends on the person and the stage they are at. Some people say lower doses feel mentally clearer, while others say going too low brings back cravings or that restless, bored feeling that can make using sound appealing again [...] & Bro, you truly only need like 4 mg a day. You're not going to get high or anything no matter how much you take. You won't be sick either. Start with 8 mg a day for a few days. Get it built up in your blood. Then take half a pill each day, 4 mg a day. You'll be fine for a long while. \hl{(uses prescriptive language)} \\ \hline
    
    How do you go about getting drug tested, while on Methadone? & Go for the drug test. Do not mention Methadone, do not bring a note, a prescription or anything. The lab will call you if they test for Methadone. Then, AND ONLY THEN, should you provide proof from your clinic. \hl{(directly answers seekers' query)} You do not want it going to anyone who has no business knowing. When they get confirmation, they will report you passed the drug test. \hl{(uses simple, non-technical language)} & Take the test and if they say anything then show the prescription. \hl{(provides limited information)} \\ 
    \bottomrule
    \end{tabular}
    \end{center}
    \caption{Sample posts, high- and low-rated comments reviewed by human annotators to infer implicit norms using community-specific signals (i.e., scores). The norms, inferred by interpreting why a comment may have received a higher or a lower rating from the community, are highlighted in gray. }
    \label{tab:implicit-norms}
\end{table*}

\subsection{LLM Judges}
As described in the main paper, we referred to existing studies~\cite{cheng2025elephantmeasuringunderstandingsocial,ziems-etal-2024-large} to develop automated LLM-based judges that determine whether a response conforms to norms. Table~\ref{tab:judge-prompt} presents the prompt used to build an LLM judge for the `use coded language' norm. A similar template was used for the other norms. 

\begin{table*}
    \centering
    \begin{tabular}{@{}p{\textwidth}@{}}
    \toprule 
    \emph{[Automated Judge]: Use Coded Language Norm} \\\\
    You are an expert annotator responsible for assessing whether a response conforms to norms surrounding support provision in an online health community (OHC). \\\\

    INSTRUCTIONS:\\
    Determine whether the response (RESPONSE) conforms to the norm (NORM) when providing support to a help-seeking query (POST) taken from an online health community. Use the definition of the norm (NORM-DEFINITION) to guide your decision. \\\\

    NORM: Use Coded Language\\\\

    NORM-DEFINITION: Use community-specific terms that seem normative in an OHC, but may not be used in offline settings. Some examples include `WD' (to refer to withdrawal), `subs' (to refer to suboxone), `oxy' (to refer to oxycodone), `takehomes', `CT' (to refer to `cold turkey'), `clean', `dirty', `abuse', `user', `users', `addict', and `addicts'. If a response uses any such terms then it conforms to the norm. \\\\

    EXAMPLES: \\
    Following are five examples, one on each line. Each example includes a RESPONSE, a binary \textsc{1} or \textsc{0} LABEL indicating whether it conforms to the norm, and a brief RATIONALE explaining the judgment.
    
    RESPONSE 1: \textsc{[positive instance text]}; LABEL: \textsc{1}; RATIONALE: \textsc{[rationale text]} \\
    RESPONSE 2: \textsc{[positive instance text]}; LABEL: \textsc{1}; RATIONALE: \textsc{[rationale text]} \\
    RESPONSE 3: \textsc{[positive instance text]}; LABEL: \textsc{1}; RATIONALE: \textsc{[rationale text]} \\
    RESPONSE 4: \textsc{[negative instance text]}; LABEL: \textsc{0}; RATIONALE: \textsc{[rationale text]} \\
    RESPONSE 5: \textsc{[negative instance text]}; LABEL: \textsc{0}; RATIONALE: \textsc{[rationale text]} \\\\

    REMEMBER: Your goal is to determine only whether the response conforms to the norm. \\\\

    INPUT FORMAT: \\
    You will receive POST: \textsc{[post text]} RESPONSE: \textsc{[response text]} \\\\

    OUTPUT FORMAT: \\
    LABEL: \textsc{1} (conforms to the norm) or \textsc{0} (does not conform to the norm). \\
    RATIONALE: \textsc{[provide rationale text]} \\\\
    
    Think step by step. Do not return anything else. \\

    \bottomrule
    \end{tabular}
    \caption{LLM judge prompt to determine whether a response conforms to the `use coded language' norm. A similar template was used for the other norms.}
    \label{tab:judge-prompt}
\end{table*}

\begin{table}[!h]
    \centering
    {
    \begin{tabular}{@{}p{0.65\columnwidth}|r|r|r@{}}
    \toprule
        \textbf{Norm} & \textbf{Pr.} & \textbf{Re.} & \textbf{F1} \\
    \midrule
        
        Avoid prescriptive guidance & 0.77 & 0.75 & 0.74 \\
        Cite sources & 0.71 & 0.70 & 0.70 \\
        Identify as a trusted expert & 0.74 & 0.75 & 0.74 \\
        Respect anonymity & 0.95 & 0.93 & 0.94 \\
        Avoid financial or product-based sourcing & 0.93 & 0.95 & 0.94 \\
        Avoid self-promotion & 0.95 & 0.93 & 0.94 \\
        Use coded language & 0.95 & 0.75 & 0.81 \\
        Use simple language & 0.93 & 0.95 & 0.94 \\
        Avoid a limited response & 0.77 & 0.75 & 0.74 \\
        Use appropriate communication strategies & 0.82 & 0.80 & 0.80 \\
        Use a dynamic style of writing & 0.91 & 0.85 & 0.86 \\
        Provide well-structured responses & 0.95 & 0.94 & 0.94 \\
        Avoid information overload & 0.93 & 0.70 & 0.76 \\
        Validate, affirm, normalize & 0.81 & 0.75 & 0.76 \\
        Instill hope & 0.95 & 0.95 & 0.95 \\
        Reflect & 0.91 & 0.85 & 0.86 \\
        Reduce isolation & 0.77 & 0.75 & 0.74 \\
        Combat stigma & 0.76 & 0.75 & 0.75 \\
        Support matching & 0.96 & 0.95 & 0.95 \\
        Personalize support & 0.86 & 0.70 & 0.72 \\
        Share lived experience & 0.85 & 0.70 & 0.71 \\
    \bottomrule
    \end{tabular}}
    \caption{Performance evaluation of LLM Judges for each norm using a human-labeled gold dataset. Pr., Re., and F1 represent precision, recall, and F1-score, respectively.}
    \label{tab:performance}
\end{table}

\subsection{Human Validation}
We performed a two-step validation of the binary labels generated by Llama-3-based automated judges. For each norm, we randomly sampled 50 posts, and corresponding GPT-4 responses, from the bigger dataset. Two authors, who are graduate students with experience in interpreting data from OHCs, manually labeled each response for conformity to a given norm. They referred to the same definitions and exemplars, which were provided to Llama-3, to guide their annotations. Next, a third annotator, with expertise in clinical substance use and social computing research, reviewed these annotations, resolved any disagreements, and provided the final gold label. These expert-vetted gold labels were then compared with the automated labels to assess validity of our automated judges. Table~\ref{tab:performance} summarizes performance evaluation for each automated judge, reporting precision, recall, and F1-score metrics, which range from 0.70 to 0.96. We further performed an error analysis to examine instances incorrectly labeled by the LLM judges (see Appendix Section ``Error Analysis'').

\subsection{Error Analysis}
Here, we briefly describe scenarios where the LLM judges erred in labeling norm conformance. We sampled responses, from the human-annotated validation dataset, that were flagged as false positives (FP) and false negatives (FN).

\paragraph{Misinterpretation of response length (FN).} LLM judge for the `avoid information overload' norm incorrectly interpreted lengthy responses as non-conforming. Human annotators indicated that the response length was not overwhelming and was, in fact, justified: either the seeker had explicitly asked for information or the context warranted greater detail. 

\paragraph{Misinterpretation of numerical values (FN).} When evaluating conformity to the `avoid prescriptive guidance' norm, the LLM judge misinterpreted numerical mentions as prescriptive guidance. For example, in an instance, the response simply echoed the dosage mentioned by the support seeker, without introducing new prescriptive guidance:

\begin{quote}
    Post: ``[...] I believe my dose was at 60, [...] now at 40 [...]'' \\
    Response: ``If you were indeed at a 60 mg dose and you've been dropped to 40 mg [...]''
\end{quote}

\paragraph{Failure to understand domain-specific terminology (FN).} In a false negative instance, the LLM judge misclassified a response combating stigma around quitting ``cold turkey,'' likely due to a lack of domain knowledge to make sense of the term~\cite{Mittal_Jung_ElSherief_Mitra_DeChoudhury_2025}.

\paragraph{Missed brief mentions (FN).} The automated judge assessing conformance to the `validate' norm occasionally failed to identify brief, early expressions of validation. For example, the following response, with a brief, introductory expression of validation, was labeled as non-conforming to the `validate' norm:

\begin{quote}
    ``It sounds like you're in a tough spot, and I can understand your frustration. (response continues)''
\end{quote}

\paragraph{Overestimating simplicity of language (FP).} When judging conformance to the `use simple language' norm, there were a few cases where the automated labels conflicted with human annotations. For example, responses with terms such as ``commendable'' or ``exacerbate,'' and phrases like ``co-occurring mental health conditions'' or ``cognitive behavioral therapy,'' without accompanying explanations, were labeled as non-conforming by humans, but conforming by the automated judge.

\paragraph{First-person references misclassified as lived experience or expertise (FP).} Sometimes, anthropomorphized outputs, with first-person references, were incorrectly flagged as conforming to `share lived experience' and `identify as a trusted expert' norms:

\begin{quote}
    ``I completely understand your frustration. [...]'' \\
    ``I'm really sorry to hear that you're going through this. [...]''
\end{quote}

\paragraph{Request for more information (FP).} According to human annotators, the automated judge incorrectly labeled a response that implicitly requested the seeker to share additional details about their history of use as conforming to the `respect anonymity' norm. Although well intentioned, human raters noted that such a response may encourage seekers to disclose personally identifiable information:

\begin{quote}
    ``[...] If you could share more details about how long you had used, that will be helpful. [...]''
\end{quote}

\end{document}